\documentclass[prd,preprint,nofootinbib]{revtex4}
\def\Journal#1#2#3#4{{#1} {\bf #2}, #3 (#4)}

\def\NPB{{\em Nucl. Phys.} {\bf B}}
\def\PLB{{\em Phys. Lett.} {\bf B}}
\def\PRL{\em Phys. Rev. Lett.}
\def\PRD{{\em Phys. Rev.} {\bf D}}

\def\PR{\em Phys. Rep.}

\def\APP{\em Astropart. Phys.}

\def\JHEP{{\em J. High Energy Phys.}}

\def\JCAP{{\em J. Cosmo. Astropart. Phys.}}
\usepackage{amsmath,graphicx}
\usepackage{latexsym}
\usepackage{ascmac}
\usepackage{amsfonts}
\usepackage{amssymb}
\usepackage{epsf}
\usepackage{bm}

\begin{document}
\title{\Large Electroweak bremsstrahlung in bino-like dark matter annihilations} 
\author{Kenta Shudo}\email{shudo@phys.cst.nihon-u.ac.jp} 
\author{Takeshi Nihei}\email{nihei@phys.cst.nihon-u.ac.jp}
  \affiliation{ 
    Department of Physics, College of Science and Technology, Nihon University, 
    1-8-14, Kanda-Surugadai, Chiyoda-ku, Tokyo, 101-8308, Japan 
   }


\baselineskip 6mm 

\begin{abstract} 
\baselineskip 6mm 
We investigate the effects of electroweak bremsstrahlung on bino-like 
neutralino dark matter pair annihilations in the minimal supersymmetric 
standard model (MSSM). 
We calculate the nonrelativistic pair annihilation cross sections via 
$W$-strahlung from leptonic final states, $\chi\chi$ $\to$ $W\ell\bar{\nu}$, 
and compare them with the contributions of the relevant two-body 
final states. 
We explore the case that sleptons lie below the TeV scale, while squarks 
are extremely heavy. 
It is found that the electroweak bremsstrahlung can give a dominant 
contribution to the cross section for some parameter regions which 
include slepton coannihilation regions with the observed relic abundance. 
We also evaluate the neutrino spectra at injection in the Sun. 
It is shown that energetic neutrinos via weak bremsstrahlung processes 
can be dominant over contributions of the two-body final states. 
\end{abstract}
\preprint{June 2013}
\vspace{1cm}
\maketitle 
%
%
\section{Introduction}
%

Clarifying the nature of cold Dark Matter (DM) is 
one of the key issues in recent 
astrophysics and cosmology~\cite{Kolb-Turner}. 
The relic abundance of cold DM in the present Universe
is determined by recent astronomical observations 
with great precision as 
$\Omega_{\chi} h^2$ $=$ 
0.1199 $\pm$ 0.0027~\cite{WMAP7}.\footnote{The density parameter 
$\Omega_{\chi}$ $=$ $\rho_{\chi}/\rho_c$ 
is the DM energy density at present normalized by
the critical density, and 
$h$ $\approx$ 0.7 is the Hubble constant in units of 
100 km/sec/Mpc.}
Among the diverse candidates, the lightest superparticle (LSP) 
in supersymmetric models is one of the most attractive ones 
for the DM particle~\cite{Jungman-etal,dm-recent-review}. 
In the minimal supersymmetric standard model (MSSM) \cite{MSSM}, 
the LSP is typically the lightest neutralino 
given by a linear combination of neutral gauginos and higgsinos 
\begin{eqnarray}
\chi & = & \chi^0_1 \ = \ 
N_{11} \tilde{B} + N_{12} \tilde{W}^3 
+ N_{13} \tilde{H}^0_1 + N_{14} \tilde{H}^0_2, 
\label{eqn:chi-1}
\end{eqnarray}
where $\tilde{B}$ is the $U(1)_Y$ gaugino (bino),  
$\tilde{W}^3$ is the neutral $SU(2)_L$ gaugino (wino), and
$\tilde{H}^0_1$ and $\tilde{H}^0_2$ are the two neutral higgsinos
with opposite hypercharges. 
The coefficients $N_{1i}$ ($i$ $=$ $1,2,3,4$) are the elements of 
the 4$\times$4 unitary matrix $N$ which diagonalizes the neutralino 
mass matrix \cite{MSSM,haber-kane}. 
Assuming the GUT relation for the gaugino masses, 
a bino-like LSP is realized for relatively light gauginos. 


Cosmic rays produced by DM annihilations in the galactic halo 
provide a way of indirect detection of DM.  
For a bino-like LSP, the dominant annihilation channel
is fermion pair production $\chi\chi$ $\to$ $f\bar{f}$. 
The neutralino 
pair annihilation cross section in the nonrelativistic limit 
is helicity suppressed ($\propto$ $m_f^2/m_{\chi}^2$) 
for light fermions 
due to the Majorana nature of the neutralino~\cite{ffbar}. 
It is known, however, that gauge boson emissions can lift 
the helicity suppression, 
since the emitted gauge boson carries unit angular momentum. 
Indeed, it has been shown that 
the bremsstrahlung $\chi\chi$ $\to$ $f\bar{f}\gamma$ 
can potentially give characteristic signals of DM 
in gamma-ray observations~\cite{bergstrom,f-fbar-gamma}. 


In recent years, the significance of electroweak bremsstrahlung 
emitting $W$/$Z$ bosons has been recognized 
in the literature~\cite{bell-etal-1st-visit,bell-etal-revisited,gamma-rays-wb,bell-cosmic-rays,bell-neutrino-osc,fukushima-etal,brems-wino-like,kumar-etal,baro-etal,antiproton,chen-kam,ew-oh2,weak-brems-misc}. 
In particular, the weak bremsstrahlung is expected to be more 
important in evaluating neutrino flux than the usual 
bremsstrahlung emitting photons, since the former emits  
primary neutrinos. 
The weak bremsstrahlung in a leptophillic dark matter model 
has been examined in Refs.~\cite{bell-etal-1st-visit,bell-etal-revisited} where
$\chi\chi$ $\to$ $W\ell\nu$ can give a dominant contribution over 
$\chi\chi$ $\to$ $\ell^+\ell^-\gamma$. 
The effect of the $W\ell\nu$ final states in this model is maximized 
in the limit where the 
dark matter mass is nearly degenerate with the mass of the 
$SU(2)_L$ doublet bosons ($\eta^0$ and $\eta^{\pm}$) 
which mediate the annihilation process similarly to sleptons. 
Gamma-ray signals in this model were investigated in 
Ref.~\cite{gamma-rays-wb}. 
Various signals, including positrons, 
were studied in Ref.~\cite{bell-cosmic-rays}. 
Neutrino spectra 
including neutrino oscillation effects 
were explored in Ref.~\cite{bell-neutrino-osc}. 
Helicity dependent effects on 
the neutrino spectra were studied 
for $SU(2)_L$ singlet Majorana fermion dark matter~\cite{fukushima-etal}. 
For a wino-like dark matter, 
initial state $W$/$Z$ radiations were found to be 
important~\cite{brems-wino-like}. 
Gamma rays from bino-like dark matter annihilations in the MSSM 
were examined including the three-body final state 
with weak bremsstrahlung~\cite{kumar-etal}. 
Neutrino signals from weak bremsstrahlung in the MSSM 
were studied for a bino-like TeV dark matter scenario~\cite{baro-etal}. 
Antiproton constraints have been studied 
in Ref.~\cite{antiproton} where effects of a longitudinal $W$-boson
emission are examined in the presence of
$SU(2)_L$ breaking effects ($m_{\eta^0}$ $\neq$ $m_{\eta^{\pm}}$), 
including the case of the constrained MSSM. 
The massive three-body final state, $Wtb$, 
was considered in Ref.~\cite{chen-kam}. 
The role of weak bremsstrahlung for the relic density of DM 
was analyzed in Ref.~\cite{ew-oh2}.

In this paper, 
we investigate effects of electroweak bremsstrahlung 
on bino-like neutralino dark matter pair annihilations 
in the phenomenological MSSM 
where various SUSY parameters are chosen freely. 
We calculate the nonrelativistic pair annihilation 
cross sections via $W$-strahlung, 
$\chi\chi$ $\to$ $W\ell\bar{\nu}$, and compare them with 
the contributions of relevant two-body final states. 
We consider the case that 
squarks are extremely heavy ($\gtrsim$ 10 TeV), while 
sleptons are much lighter than squarks. 
In this case, the weak bremsstrahlungs with quarks, 
$\chi\chi$ $\to$ $Wud$, $Wcs$, $Wtb$ via $t$- and $u$-channel 
squark exchanges are suppressed. 
Then we discuss leptonic processes with primary neutrinos 
\begin{eqnarray} 
\chi\chi  & \to & W^+ \ell \bar{\nu}_{\ell} \, + \, {\rm h.c.}
 \ \ \ \ (\,\ell \, = \, e, \, \mu, \, \tau \,). 
\label{eqn:chichi-to-Wlnu}
\end{eqnarray}

This paper is organized as follows. 
In Sec.~\ref{sec:interactions}, we describe 
the relevant MSSM interactions. 
%
In Sec.~\ref{sec:cross-section}, 
we calculate the cross sections and neutrino spectra 
for the weak bremsstrahlung.
In Sec.~\ref{sec:NWA}, 
we discuss the narrow width approximation for the 
massive weak bremsstrahlung process $\chi\chi$ $\to$ $Wtb$. 
In Sec.~\ref{sec:numerical-results}, we present our numerical results. 
Finally concluding remarks are given in Sec.~\ref{sec:conclusions}. 
A simplified expression for the cross section 
in the unbroken $SU(2)_L$ limit 
in the slepton sector is provided in Appendix~\ref{app:toy-model}. 
The contributions of two-body final states are 
listed in Appendix~\ref{app:2-body}. 
%

%
\section{Relevant MSSM interactions}
\label{sec:interactions}
%

The relevant interaction Lagrangian of 
the neutralino $\chi$ with leptons and sleptons 
can be written as 
\begin{eqnarray}
{\cal L}_{\chi \ell \tilde{\ell}}
 & = & 
C^{(\nu)}_L 
\sum_{\ell} 
\overline{\chi} P_L \nu_{\ell} \tilde{\nu}^*_{\ell}
 + 
\sum_{\ell} \sum_{I=1,2}
\overline{\chi} \left(  
C_{LI}^{(\ell)} P_L  
+ C_{RI}^{(\ell)} P_R \right) \ell \tilde{\ell}^*_I
 + {\rm h.c.}, 
\label{eqn:Lag-chi-f-sf}
\end{eqnarray}
where 
$P_L$ $=$ $\frac{1-\gamma_5}{2}$ and $P_R$ $=$ $\frac{1+\gamma_5}{2}$. 
The fields $\tilde{\ell}_I$ ($I$ $=$ 1, 2) denote 
the charged slepton mass eigenstates, and $\tilde{\nu}_{\ell}$
is the sneutrino ($\ell$ $=$ $e$, $\mu$, $\tau$). 
Flavor mixings and CP violations are neglected. 
The coupling constants in Eq.~(\ref{eqn:Lag-chi-f-sf}) 
are given by 
\begin{eqnarray}
C^{(\nu)}_L & = & 
\frac{1}{\sqrt{2}} \left( -\, gN_{12} + g' N_{11} \right), \nonumber \\
C^{(\ell)}_{LI}  & = & 
\frac{1}{\sqrt{2}} \left( gN_{12} + g' N_{11} \right) 
\big( \widetilde{V}_{\ell} \big)_{I1} 
- h_{\ell} N_{13} \big( \widetilde{V}_{\ell} \big)_{I2}, 
\label{eqn:C-fL-I} \\
C^{(\ell)}_{RI}  & = & 
-\, h_{\ell} N_{13} \big( \widetilde{V}_{\ell} \big)_{I1} 
- \sqrt{2} g' N_{11} \big( \widetilde{V}_{\ell} \big)_{I2}, \nonumber 
\end{eqnarray}
where $g'$ and $g$ denote the gauge coupling constant 
for the $U(1)_Y$ and $SU(2)_L$, 
$h_{\ell}$ $=$ $gm_{\ell}/(\sqrt{2}m_W \cos\beta)$ 
is the Yukawa coupling constant for the lepton, and 
$N_{ij}$ is the element of the unitary matrix 
to diagonalize the neutralino mass matrix \cite{haber-kane,NRR2}. 
The vacuum angle $\beta$ is given by $\tan \beta$ $=$ $v_2/v_1$, 
where $v_1$ and $v_2$ are the vacuum expectation values  
of the two neutral Higgs bosons. 
The unitary matrix $\widetilde{V}_{\ell}$ diagonalizes 
the charged slepton mass squared matrix as 
$\widetilde{V}_{\ell}  M_{\tilde{\ell}}^2 \widetilde{V}_{\ell}^{\dagger}$ 
$=$ diag ($m_{\tilde{\ell}1}^2$, $m_{\tilde{\ell}2}^2$). 
The mass squared matrix, neglecting the $m_{\ell}^2$ terms, 
is approximately given by~\cite{haber-kane} 
\begin{eqnarray}
M_{\tilde{\ell}}^2 & = &
\left(
\begin{array}{cc}
m_{\tilde{\ell}L}^2 + m_Z^2 
\left(-\,{\textstyle \frac{1}{2}}+s_W^2 \right) \cos 2\beta 
 & m_{\ell}(A_{\ell}-\mu\tan\beta)  \\
m_{\ell}(A_{\ell}-\mu\tan\beta) 
 & m_{\tilde{\ell}R}^2 - m_Z^2 s_W^2 \cos 2\beta \\
\end{array}
\right),
\label{eqn:slepton-m2}
\end{eqnarray}
where $m_{\tilde{\ell}L}^2$ and $m_{\tilde{\ell}R}^2$ 
are the soft supersymmetry (SUSY) breaking mass 
parameters for the left- and right-handed sleptons, respectively, 
$s_W$ $=$ $\sin \theta_W$, 
and $A_{\ell}$ is the trilinear scalar coupling constant 
for the slepton. 
The mass eigenstates are related with the chiral bases 
$\tilde{\ell}_{L}$ and $\tilde{\ell}_{R}$ 
as $\tilde{\ell}_I$ $=$ 
$\big( \widetilde{V}_{\ell} \big)_{I1} \tilde{\ell}_{L}$
$+$ $\big( \widetilde{V}_{\ell} \big)_{I2} \tilde{\ell}_{R}$. 
The sneutrino mass squared is given by 
$m_{\tilde{\nu}}^2$ $=$ 
$m_{\tilde{\ell}L}^2 + \frac{1}{2}m_Z^2 \cos 2\beta $. 
The $W$-boson emission from the slepton 
involves the following interaction:  
\begin{eqnarray} 
{\cal L}_{W\tilde{\ell}\tilde{\nu}}
 & = & 
-\, \sum_{\ell} \sum_{I=1,2} \frac{ig}{\sqrt{2}} (V_{\tilde{\ell}})_{I1} 
\Big( \tilde{\ell}_I^* \stackrel{\leftrightarrow}{\partial^{\mu}} 
 \tilde{\nu} \Big) W_{\mu}^- 
 + {\rm h.c.}
\end{eqnarray}
We follow the convention of Ref.~\cite{haber-kane} for 
the MSSM parameters.

We assume the GUT relation for the gaugino masses: 
$M_1$ $=$ $\frac{5}{3} M_2 \tan^2 \theta_W$ 
where $M_2$ and $M_1$ are the gaugino masses for the 
$SU(2)_L$ and $U(1)_Y$ gauginos. 
Numerically, this implies $M_1$ $\sim$ $M_2/2$. 
The neutralino in Eq.~(\ref{eqn:chi-1}) is 
bino-like for $M_1$ $\ll$ $|\mu|$, 
where $\mu$ is the Higgsino mass parameter. 
%

%
%
\section{Cross sections for weak bremsstrahlung}
\label{sec:cross-section}
%

In this section, we present the cross sections 
for the weak bremsstrahlung process (\ref{eqn:chichi-to-Wlnu}) 
in the nonrelativistic limit $v$ $\to$ 0, 
where $v$ is the relative velocity between the two 
neutralinos in the center of mass frame. 

\begin{figure}[t]
\includegraphics[width=15cm]{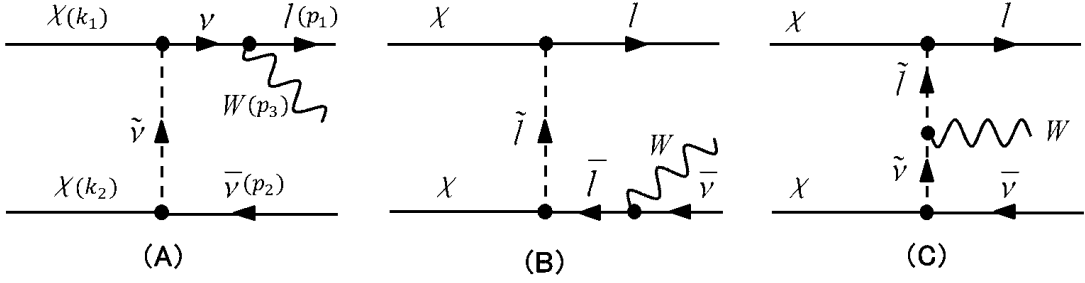}
\caption[cap:diagrams]{
\baselineskip 6mm 
Feynman diagrams for the weak bremsstrahlung process
$\chi\chi$ $\to$ $W^+ \overline{\nu}_{\ell} \ell $. 
The $W$ boson is emitted (A) from the virtual neutrino, 
(B) from the virtual charged lepton, and 
(C) from the virtual slepton line. 
Corresponding $u$-channel diagrams are not shown. 
}
\label{fig:diagrams}
\end{figure}

The relevant Feynman diagrams 
for $\chi\chi$ $\to$ $W^+ \ell \bar{\nu}_{\ell}$ 
via $t$- and $u$-channel 
slepton exchange are shown in Fig.~\ref{fig:diagrams}. 
The diagram A (B) proceeds via pair production 
$\chi\chi$ $\to$ $\nu^*\bar{\nu}$ 
($\chi\chi$ $\to$ $\ell\bar{\ell}^*$) 
followed by the $W$-boson emission from the neutrino (charged lepton). 
In the diagram C, the $W$ boson is emitted by the virtual 
sleptons. 

There exist other diagrams which can 
contribute to the process (\ref{eqn:chichi-to-Wlnu}) 
in principle. 
However, in the parameter region we consider, they
give only negligible effects. 
Contributions via 
$s$-channel $Z$-boson and pseudoscalar Higgs ($A$) exchange 
$\chi\chi$ $\to$ $Z^*$/$A^*$ $\to$ $\ell \bar{\ell}^*$
$\to$ $W\ell\bar{\nu}$ give no significant effects, 
since the neutralino coupling to the $Z$ boson or 
the pseudoscalar Higgs boson is highly suppressed for
a bino-like LSP.\footnote{Note, however, that a $s$-channel diagram is relevant for $t\bar{t}$ and $b\bar{b}$ final states, since the $t$- and $u$-channel squark exchange diagrams are suppressed for extremely heavy squarks.} 
Initial state radiation, in which an initial neutralino 
emits the $W$ boson, is negligible for a bino-like LSP, 
since a pure bino does not couple to a $W$ boson. 
Contributions via $W$-boson pair production 
followed by the $W$-boson decay to the leptonic pair, 
$\chi\chi$ $\to$ $W W^*$ $\to$ $W \ell \bar{\nu}$, 
are negligible due to the suppressed coupling of 
the bino-like LSP to the $W$ boson.\footnote{This contribution is taken into account as a part of the two-body process $\chi\chi$ $\to$ $WW$ in Sec.~\ref{sec:numerical-results}.}  

The 4-momentum of each particle in Fig.~\ref{fig:diagrams} 
is assigned as follows:  
\begin{eqnarray}
\chi (k_1) + \chi (k_2) 
 & \to & \ell (p_1) + \overline{\nu_\ell} (p_2) + W^+(p_3). 
\end{eqnarray}
%
%
We define the sum of the $t$- and $u$-channel diagram for 
each diagram A, B and C in Fig.~\ref{fig:diagrams} as 
${\cal M}_{\rm A}$ $=$ ${\cal M}_{{\rm A}t}$ $+$ ${\cal M}_{{\rm A}u}$, 
${\cal M}_{\rm B}$ $=$ ${\cal M}_{{\rm B}t}$ $+$ ${\cal M}_{{\rm B}u}$ 
and 
${\cal M}_{\rm C}$ $=$ ${\cal M}_{{\rm C}t}$ $+$ ${\cal M}_{{\rm C}u}$, 
respectively. 
After the Fierz rearrangements, 
these matrix elements can be written as 
{\allowdisplaybreaks
\begin{eqnarray} 
{\cal M}_{\rm A} 
 & = & 
\frac{ig}{2\sqrt{2}} \frac{1}{t_2-m^2_{\tilde{\nu}}}
\frac{1}{q_1^2} 
( C^{(\nu)}_L )^2 
\big( \overline{u_{\ell}} \!\not{\! \varepsilon}^{\,*}_3 
 \!\not{\! q}_1  \gamma^{\alpha} P_L v_{\nu} \big) 
\big( \overline{v_{\chi}} \gamma_{\alpha} \gamma_5 u_{\chi} \big), 
\label{eqn:amp-MA} \\ 
{\cal M}_{\rm B} 
 & = & 
\sum_{I=1,2} 
\frac{ig}{2\sqrt{2}} \frac{1}{t_1-m^2_{\tilde{\ell}I}}
\frac{1}{q_2^2} C^{(\ell)}_{LI}  \nonumber \\
 & & 
\hspace{1mm}
\times \, 
\bigg[ \, 
-\, C^{(\ell)}_{LI} 
\big( \overline{u_{\ell}} \gamma^{\alpha}
 \!\not{\! q}_2 \!\not{\! \varepsilon}^{\,*}_3 P_L v_{\nu} \big) 
\big( \overline{v_{\chi}} \gamma_{\alpha} \gamma_5 u_{\chi} \big) 
+ \, C^{(\ell)}_{RI}
\big( \overline{u_{\ell}} 
 \!\not{\! q}_2 \!\not{\! \varepsilon}^{\,*}_3 P_L v_{\nu} \big) 
\big( \overline{v_{\chi}} \gamma_5 u_{\chi} \big) 
 \ \bigg], 
\label{eqn:amp-MB} \\ 
{\cal M}_{\rm C} 
 & = & 
\sum_{I=1,2} 
\frac{ig}{2\sqrt{2}} 
\frac{1}{t_2-m^2_{\tilde{\nu}}} \frac{1}{t_1-m^2_{\tilde{\ell}I}}
(V_{\tilde{\ell}})_{I1} 
 \left[ \varepsilon^{\,*}_3 \cdot (p_1-p_2) \right] 
C^{(\nu)}_L \nonumber \\
 & & 
\hspace{3mm}
\times \bigg[ \ C^{(\ell)}_{LI} 
\big( \overline{u_{\ell}} \gamma^{\alpha} P_L v_{\nu} \big) 
\big( \overline{v_{\chi}} \gamma_{\alpha} \gamma_5 u_{\chi} \big)
- \, C^{(\ell)}_{RI}
\big( \overline{u_{\ell}} P_L v_{\nu} \big) 
\big( \overline{v_{\chi}} \gamma_5 u_{\chi} \big) 
 \ \bigg], 
\label{eqn:amp-MC} 
\end{eqnarray}
}
where $q_1$ $=$ $p_1+p_3$, $q_2$ $=$ $p_2+p_3$, 
and $\varepsilon_3$ $=$ $\varepsilon (p_3)$ is the 
polarization vector of the $W$ boson. 
The spinors $u$ and $v$ are denoted as 
$u_{\chi}$ $=$ $u(k_1)$, $v_{\chi}$ $=$ $v(k_2)$, 
$u_{\ell}$ $=$ $u(p_1)$, $v_{\nu}$ $=$ $v(p_2)$, 
where the spin indices are suppressed. 
The Lorentz invariants $t_1$ and $t_2$ are defined as 
\begin{eqnarray}
t_1 & = & (k_1-p_1)^2 \ = \ (k_2-p_1)^2 \ = \ u_1,  \nonumber \\
t_2 & = & (k_1-p_2)^2 \ = \ (k_2-p_2)^2 \ = \ u_2. 
\label{eqn:t1-t2}
\end{eqnarray}
Note that $k_1$ $=$ $k_2$ 
in the nonrelativistic limit $v$ $\to$ 0. 
In evaluating the matrix elements, 
we neglect the lepton masses compared with $m_W$ and $m_{\chi}$, 
while we take into account the lepton masses 
in slepton mass matrices to keep potentially large left-right mixings 
for the sleptons.\footnote{In the case of electromagnetic bremsstrahlung $\chi\chi$ $\to$ $\ell^+\ell^-\gamma$, this leads to an infrared divergence in the total cross section, which should be cancelled by virtual one-loop corrections to the two-body annihilation rate. In the case of $W$-strahlung, however, the total cross section is free from an infrared divergence even when the lepton mass is neglected, since the denominator of the lepton propagators in Eqs. (\ref{eqn:amp-MA}) and (\ref{eqn:amp-MB}) does not vanish due to $q_i^2$ $\geq$ $m_W^2$ ($i$ $=$ 1, 2).}
The Gordon decomposition for the s-wave limit, 
$m_{\chi} \overline{v_{\chi}} \gamma^{\alpha} \gamma_5 u_{\chi}$ 
$=$ $-\, k_1^{\alpha} \overline{v_{\chi}} \gamma_5 u_{\chi}$, 
can be used to further simplify 
Eqs.~(\ref{eqn:amp-MA})-(\ref{eqn:amp-MC}).

The differential cross section for the process 
$\chi\chi$ $\to$ $W^+ \ell \bar{\nu}_{\ell}$ is given by 
\begin{eqnarray}
\frac{d^2 (\sigma v)_{W\ell\bar{\nu}}}{dE_W \, dE_{\nu}} 
 & = & 
\frac{1}{512\pi^3 m_{\chi}^2} \sum_{\rm spins} 
 |{\cal M}_{\rm A}+{\cal M}_{\rm B}+{\cal M}_{\rm C}|^2, 
\label{eqn:diff-cross-3}
\end{eqnarray}
where $E_W$ and $E_{\nu}$ are the energy of the $W$ boson and 
the neutrino at the center of mass frame. 
To include the charge-conjugated process as in 
Eq.~(\ref{eqn:chichi-to-Wlnu}), 
the differential cross section in 
Eq.~(\ref{eqn:diff-cross-3}) must be doubled. 
The helicity sum of the matrix element squared for diagrams 
A, B and C are given as follows: 
{\allowdisplaybreaks
\begin{eqnarray}
\sum_{\rm spins} \left| {\cal M}_{\rm A} \right|^2 
 & = & 
\frac{g^2}{4m_W^2}
\frac{1}{(t_2-m^2_{\tilde{\nu}})^2} ( C^{(\nu)}_L )^4 F_1, 
\label{eqn:M2AA} \\ 
%
\sum_{\rm spins} \left| {\cal M}_{\rm B} \right|^2 
 & = & 
\sum_{I,J}
\frac{g^2}{4m_W^2} 
\frac{1}{t_1-m^2_{\tilde{\ell}I}} \frac{1}{t_1-m^2_{\tilde{\ell}J}} 
    \nonumber \\
 & & \hspace{5mm} \times \, 
C^{(\ell)}_{LI} C^{(\ell)}_{LJ}
\bigg[ \, 
C^{(\ell)}_{LI} C^{(\ell)}_{LJ} F_1 
 + \frac{4m_{\chi}^2  C^{(\ell)}_{RI} C^{(\ell)}_{RJ}}{z^2} F_2 \, \bigg], 
\label{eqn:M2BB} \\ 
%
\sum_{\rm spins} \left| {\cal M}_{\rm C} \right|^2 
 & = & 
\sum_{I,J}
\frac{g^2}{16 m_W^2} 
\frac{1}{(t_2-m^2_{\tilde{\nu}})^2} 
\frac{1}{t_1-m^2_{\tilde{\ell}I}} \frac{1}{t_1-m^2_{\tilde{\ell}J}} 
(V_{\tilde{\ell}})_{I1} (V_{\tilde{\ell}})_{J1} ( C^{(\nu)}_L )^2 
  \nonumber \\
 & & \hspace{0mm} \times \, 
\big[ (x-z)^2+4m_W^2 y \big] 
\bigg[ \, C^{(\ell)}_{LI} C^{(\ell)}_{LJ} F_3  
+ C^{(\ell)}_{RI} C^{(\ell)}_{RJ} \cdot 4m_{\chi}^2 y \, \bigg], 
\label{eqn:M2CC}  \\
%
\sum_{\rm spins} {\cal M}_{\rm A} {\cal M}_{\rm B}^* + {\rm h.c.} 
 & = &
-\, \sum_{I}
\frac{g^2}{2m_W^2} 
\frac{1}{t_2-m^2_{\tilde{\nu}}} \frac{1}{t_1-m^2_{\tilde{\ell}I}} 
( C^{(\nu)}_L )^2 (C^{(\ell)}_{LI})^2  F_1, 
\label{eqn:M2AB} \\ 
%
\sum_{\rm spins} {\cal M}_{\rm A} {\cal M}_{\rm C}^* + {\rm h.c.} 
 & = &
\sum_{I}
\frac{g^2}{4m_W^2} 
\frac{1}{(t_2-m^2_{\tilde{\nu}})^2}  
\frac{1}{t_1-m^2_{\tilde{\ell}I}} (V_{\tilde{\ell}})_{I1} 
(C^{(\nu)}_L)^3 C^{(\ell)}_{LI} F_4, 
\label{eqn:M2AC} \\ 
%
\sum_{\rm spins} {\cal M}_{\rm B} {\cal M}_{\rm C}^* + {\rm h.c.} 
 & = &
-\, \sum_{I,J} \frac{g^2}{4m_W^2} 
\frac{1}{t_2-m^2_{\tilde{\nu}}} 
\frac{1}{t_1-m^2_{\tilde{\ell}I}} \frac{1}{t_1-m^2_{\tilde{\ell}J}} 
(V_{\tilde{\ell}})_{I1} C^{(\nu)}_L C^{(\ell)}_{LJ}  
  \nonumber \\
 & & \hspace{3.3cm} \times 
\bigg[ 
C^{(\ell)}_{LI} C^{(\ell)}_{LJ} F_4 
 + \ \frac{4m_{\chi}^2}{z} C^{(\ell)}_{RI} C^{(\ell)}_{RJ} F_5 
\bigg].  
\label{eqn:M2BC} 
\end{eqnarray}
The auxiliary functions $F_1$, $F_2$, $\cdots$, $F_5$ are 
given by 
{\allowdisplaybreaks
\begin{eqnarray}
F_1 & = & xz+2m_W^2(y-2m_{\chi}^2), \nonumber \\ 
F_2 & = & 
z(y-2m_W^2)(z-m_W^2) 
- m_W^2 \Big[ z (y-8m_{\chi}^2) + 8m_W^2 m_{\chi}^2 \Big], 
   \nonumber \\ 
F_3 & = & x F_1 -2m_W^2 y, \\
F_4 & = & (x-z) F_3, \nonumber \\ 
F_5 & = & 
y \Big[ z (x-z)  + 2 m_W^2 (z-4m_{\chi}^2) \Big], \nonumber 
\end{eqnarray}
} 
where the Lorentz invariants $x$, $y$ and $z$ are defined by 
\begin{eqnarray}
x & = & (p_1+p_3)^2, \ \ \ \ \ 
y \ = \ (p_1+p_2)^2, \ \ \ \ \ 
z \ = \ (p_2+p_3)^2. 
\label{eqn:xyz}
\end{eqnarray}
Note that $x$ $+$ $y$ $+$ $z$ $=$ $4m_{\chi}^2 + m_W^2$, 
where $m_{\chi}$ and $m_W$ 
denote the mass of the neutralino and $W$ boson, respectively. 
The Lorentz invariants in Eqs.~(\ref{eqn:t1-t2}) 
and (\ref{eqn:xyz}) can be written in terms of the
energies as 
$x$ $=$ $4m_{\chi}(m_{\chi}-E_{\nu})$, 
$y$ $=$ $4m_{\chi}(m_{\chi}-E_W)$ $+$ $m_W^2$, 
$z$ $=$ $4m_{\chi}(m_{\chi}-E_{\ell})$, 
$t_1$ $=$ $\frac{z}{2}$ $-$ $m_{\chi}^2$, 
$t_2$ $=$ $\frac{x}{2}$ $-$ $m_{\chi}^2$, 
where $E_{\ell}$ is the energy of the charged lepton at the 
center of mass frame of the initial neutralinos, 
and the lepton mass is neglected.

Let us discuss the behavior in the heavy slepton limit: 
$m_{\tilde{\ell}I}$ $=$ $m_{\tilde{\nu}}$ $\equiv$ 
$\tilde{m}$ $\to$ $\infty$. 
The slepton mass dependence of each amplitude is  
${\cal M}_{\rm A}$, ${\cal M}_{\rm B}$ $\sim$ $1/\tilde{m}^2$, 
and ${\cal M}_{\rm C}$ $\sim$ $1/\tilde{m}^4$. 
However, by summing up Eqs.~(\ref{eqn:M2AA}), 
(\ref{eqn:M2BB}) and (\ref{eqn:M2AB}), 
the leading $1/\tilde{m}^2$ terms cancel out between diagrams 
A and B, resulting in the total amplitude suppressed as
$\sim$ $1/\tilde{m}^4$~\cite{bell-etal-revisited}. 
On the other hand, the amplitude for the leptonic 
two-body process, $\chi\chi$ $\to$ $\tau^+\tau^-$, 
is suppressed only by $\sim$ $1/\tilde{m}^2$. 
Therefore, 
the ratio $(\sigma v)_{W\tau\nu}/(\sigma v)_{\tau^+\tau^-}$ 
falls down for heavy sleptons.


We also evaluate neutrino spectra at injection from 
the center of the Sun. 
The primary neutrino spectrum via $\chi\chi$ $\to$ $W\ell \bar{\nu}$ 
is obtained by integrating Eq.~(\ref{eqn:diff-cross-3}) over $E_W$ as 
\begin{eqnarray}
\frac{d(\sigma v)_{W\ell\bar{\nu}}}{dE_{\nu}}
 & = & 
\int_{E_{W}^{\rm min} (E_{\nu})}^{E_{W}^{\rm max}} 
dE_W \frac{d^2 (\sigma v)_{W\ell\bar{\nu}}}{dE_W dE_{\nu}}, 
\label{eqn:dndE-primary} 
\end{eqnarray} 
where 
\begin{eqnarray} 
E_{W}^{\rm min} (E_{\nu})
 & = & 
m_{\chi}-E_{\nu} + \frac{m_W^2}{4(m_{\chi}-E_{\nu})}, 
 \nonumber \\
E_{W}^{\rm max} & = & 
m_{\chi} \left( 1 + \frac{m_W^2}{4 m_{\chi}^2} \right). 
\end{eqnarray} 
The cross section is obtained by 
integrating Eq.~(\ref{eqn:dndE-primary}) over $E_{\nu}$ 
in the range  
0 $<$ $E_{\nu}$ $<$ 
$m_{\chi}\left( 1-\frac{m_W^2}{4 m_{\chi}^2}\right)$. 
This integration can be done analytically, although 
the expressions are lengthy.

The secondary neutrino spectra from decay of 
the $W$ boson and the tau lepton are evaluated as follows. 
The neutrino spectrum via $W$-boson decay is written as 
\begin{eqnarray}
\left. 
\frac{d (\sigma v)_{W\ell\bar{\nu}}}{dE_{\nu}} \right|_{{\rm from } \, W} 
 & = & 
\int_{m_W}^{E_{W}^{\rm max}} dE_W \frac{d (\sigma v)_{W\ell\bar{\nu}}}{dE_W} 
  \left[ \left( \frac{dN_{\nu}}{dE_{\nu}} \right)_W (E_W, E_{\nu}) \right], 
\label{eqn:spectrum-3body}
\end{eqnarray} 
where 
$\left( \frac{dN_{\nu}}{dE_{\nu}} \right)_W (E_W, E_{\nu})$ 
is the neutrino energy distribution per $W$-boson decay with energy $E_W$, 
and $d (\sigma v)_{W\ell\bar{\nu}}/dE_W$ is the $W$-boson 
spectrum obtained by integrating Eq.~(\ref{eqn:diff-cross-3}) 
over $E_{\nu}$.
The secondary neutrino spectrum from tau decay 
is obtained in a similar fashion using 
the neutrino distribution per tau decay, 
$\left( \frac{dN_{\nu}}{dE_{\nu}} \right)_{\tau} (E_{\tau}, E_{\nu})$. 
In evaluating the contributions via relevant two-body processes 
$\chi\chi$ $\to$ $\tau^+\tau^-$, $t\bar{t}$, $b\bar{b}$ 
and $W^+W^-$, we further need the neutrino distributions from 
the top and bottom quarks. 
Neutrino distribution 
$\left( \frac{dN_{\nu}}{dE_{\nu}} \right)_i (E_i, E_{\nu})$ 
from the parent particle ($i$ $=$ $W$, $\tau$, $t$, $b$) 
with energy $E_i$ is affected by matter effects in the Sun. 
We neglect decay of the light quarks and the muon, 
since they stop before decay in the Sun. 
We also neglect the contribution of the charm quark, 
since it is subdominant compared with that of the bottom quark. 
For the energy distributions  
$\left( \frac{dN_{\nu}}{dE_{\nu}} \right)_i$ in the Sun, 
we use the result of Ref.~\cite{secondary-neutrino} 
where the distributions are obtained with 
the Monte Carlo code PYTHIA~\cite{pythia}. 
For the energy of the parent particles 
not tabulated in the reference, 
we simply adopt linear interpolations.

%
\section{Narrow width approximation for $\bm{\chi \chi \to Wtb}$} 
\label{sec:NWA}


In this section, we discuss the relation between the 
massive weak bremsstrahlung process $\chi\chi$ $\to$ $Wtb$ 
and the corresponding two-body process $\chi\chi$ $\to$ $t\bar{t}$
using a narrow width approximation.

The cross section for $Wtb$ can be obtained 
from ${\cal M}_{\rm A}$, ${\cal M}_{\rm B}$ and ${\cal M}_{\rm C}$ 
by replacing $\nu$ and $\ell$ with $t$ and $b$, respectively, 
and taking both left- and right-handed stops into account. 
When $m_{\chi}$ $>$ $m_t$, the top quark pair production 
$\chi\chi$ $\to$ $t\bar{t}$ opens up. 
Then, the three-body cross section for  
$\chi\chi$ $\to$ $Wtb$ calculated with only ${\cal M}_{\rm A}$ 
reduces to the cross section of the two-body process 
$\chi\chi$ $\to$ $t\bar{t}$ 
evaluated with only $t$- and $u$-channel diagrams. 
In this sense, 
the massive weak bremsstrahlung 
$\chi\chi$ $\to$ $W^+ \bar{t} b$ is included in the 
two-body process $\chi\chi$ $\to$ $t\bar{t}$ followed by 
the on-shell top quark decay for 
$m_{\chi}$ $>$ $m_t$~\cite{chen-kam}. 
On the other hand, the leptonic $W$-strahlung in 
Eq.~(\ref{eqn:chichi-to-Wlnu}) can never be included 
in any two-body process, since a $W$-boson emission from on-shell
lepton is kinematically forbidden.

We have calculated the cross section for the $Wtb$ final state 
in the same way as the leptonic case. 
Integrating the differential cross section 
$d^2 (\sigma v)_{W\bar{t}b}^{(\tilde{t})}/(dE_W dE_t)$ over $E_W$, 
the top quark energy distribution via only stop exchange 
can be obtained as 
\begin{eqnarray}
\frac{d (\sigma v)_{W\bar{t}b}^{(\tilde{t})}}{dE_t} 
 & = & 
\sum_{I,J}
\frac{N_c g^2}{512\pi^3 m_W^2} \sqrt{E_t^2-m_t^2}
\frac{\Delta_t + m_t^2+2m_W^2}{\Delta_t^2+\Gamma_t^2m_t^2}
  \nonumber \\
 & & \ \ \ 
\times \, \frac{\left( \Delta_t + m_t^2 -m_W^2 \right)^2}{\Delta_t+m_t^2}
\frac{1}{t_2-m^2_{\tilde{t}I}}\frac{1}{t_2-m^2_{\tilde{t}J}} 
  \nonumber \\
 & & \ \ \ \times \, 
\bigg[ 
(C^{(t)}_{LI})^2 (C^{(t)}_{LJ})^2 f_t (E_t)
+ m_t m_{\chi}\Big( (C^{(t)}_{LI})^2 C^{(t)}_{RJ} D^{(t)}_{LJ} 
                   + (C^{(t)}_{LJ})^2 C^{(t)}_{RI} D^{(t)}_{LI} \Big) 
  \nonumber \\
 & & \hspace{6cm}
+ \ \frac{m_{\chi}^2f_t(E_t)}{\Delta_t+m_t^2} 
               C^{(t)}_{RI} D^{(t)}_{LI} C^{(t)}_{RJ} D^{(t)}_{LJ} \bigg], 
\label{eqn:dsvdEt}
\end{eqnarray}
where $\Delta_t$ $=$ $x-m_t^2$ $=$ 
$4m_{\chi} (m_{\chi}-E_t)$, 
$\Gamma_t$ is the decay width for the top quark, 
$N_c$ $=$ 3 is the color factor, and 
\begin{eqnarray}
f_t (E_t) & = & 2E_t(m_{\chi}-E_t) + m_t^2. 
\end{eqnarray}
The coupling constants $C^{(t)}_{LI}$ and $C^{(t)}_{RI}$ can be defined 
in a similar fashion as $C^{(\ell)}_{LI}$ and $C^{(\ell)}_{RI}$ 
in Eq.~(\ref{eqn:C-fL-I}). 
The constant $D^{(t)}_{LI}$ is defined by 
\begin{eqnarray}
m_{\chi} D^{(t)}_{LI} & = & 2m_{\chi} C^{(t)}_{LI} + m_t C^{(t)}_{RI}. 
\end{eqnarray}
Integration of Eq.~(\ref{eqn:dsvdEt}) over $E_t$ gives 
the cross section $(\sigma v)_{W\bar{t}b}^{(\tilde{t})}$. 
If the integration range includes the pole $E_t$ $=$ $m_{\chi}$
($x$ $=$ $m_t^2$), the narrow width approximation can be 
justified where the propagator of the top quark can be 
replaced with the delta function as 
\begin{eqnarray}
\frac{1}{(x-m_t^2)^2+\Gamma_t^2m_t^2} 
 & \approx & \frac{\pi}{\Gamma_t m_t} \delta (x-m_t^2). 
\end{eqnarray}
Under this approximation, the integration of 
Eq.~(\ref{eqn:dsvdEt}) over $x$ (equivalently over $E_t$) 
indeed reduces to 
the two-body s-wave cross section 
via $t$- and $u$-channel stop exchange~\cite{Jungman-etal} 
(see Appendix~\ref{app:2-body}) 
\begin{eqnarray}
(\sigma v)_{t\bar{t}}^{(\tilde{t})} 
& = & 
\frac{N_c}{32\pi}
\sqrt{1-\frac{m_t^2}{m_{\chi}^2}} 
\left| 
\sum_{I=1,2} 
\frac{m_t \Big( (C^{(t)}_{LI})^2 + (C^{(t)}_{RI})^2 \Big) + 2m_{\chi}C^{(t)}_{LI} C^{(t)}_{RI}}{m_t^2 - m_{\chi}^2 - m_{\tilde{t}I}^2} 
\right|^2, 
\label{eqn:ttbar-TandU}
\end{eqnarray}
where we use the expression for the top-quark decay width \cite{top-decay}
\begin{eqnarray}
\Gamma_t 
 & = & 
\frac{g^2}{64 \pi} \frac{m_t^3}{m_W^2}
\left( 1-\frac{m_W^2}{m_t^2} \right)^2 
\left( 1 + 2\frac{m_W^2}{m_t^2} \right). 
\end{eqnarray}

We neglect the s-channel contributions for the three-body processes 
for simplicity. 
This is a good approximation for the leptonic process 
$\chi\chi$ $\to$ $W\ell\nu$. 
For the $Wtb$ final state, however, this is not a good 
approximation, since squarks are extremely heavy in the 
present analysis. 
In our numerical calculation, we use the total two-body 
expression for $\chi\chi$ $\to$ $t\bar{t}$ in Eq.~(\ref{eqn:sv-ffbar})
rather than in Eq.~(\ref{eqn:ttbar-TandU}). 
Therefore, our result for the $Wtb$ final state 
below the $t\bar{t}$ threshold $m_{\chi}$ $\lesssim$ $m_t$ 
is not smoothly connected to that for the $t\bar{t}$ final state. 
This does not affect our conclusion, since the $Wtb$ contribution 
below the threshold is subdominant in the parameter range 
we consider in the present analysis.

%
\section{Numerical results}
\label{sec:numerical-results}

In this section, we present our numerical results. 
In this analysis, we examine the phenomenological MSSM, 
where various SUSY parameters are chosen freely. 
In order to find the parameter ranges where 
the weak bremsstrahlung is important, 
we consider the scenario 
that the sleptons lie below the TeV scale, 
while the squarks are extremely heavy. 
Throughout the analyses, 
squark mass parameters are taken as 
$m_{\tilde{q}}$ $\gtrsim$ 10 TeV for all the squarks 
to be consistent with the null results of superparticle 
searches at the LHC~\cite{LHC-susy-constraint}. 
The pseudoscalar Higgs boson mass is fixed at 
$m_A$ $=$ 2 TeV in every figure. 
For the slepton mass parameters, 
we assume a common value 
$m_{\tilde{e}}$ $=$ $m_{\tilde{\mu}}$ for the 
left- and right-handed soft SUSY breaking masses for 
the selectrons and smuons, while 
the left- and right-handed stau mass parameters, 
$m_{\tilde{\tau}L}$ and $m_{\tilde{\tau}R}$, 
are chosen independently. 
For the trilinear scalar couplings, 
we vary $A_t$, $A_b$ and $A_{\tau}$ for the stop, sbottom and stau, 
respectively. 
All the others are set to zero: 
$A_q$ $=$ 0 ($q$ $\neq$ $t$, $b$) for the other squarks, 
and $A_e$ $=$ $A_{\mu}$ $=$ 0.


The s-wave cross section times the relative velocity, $\sigma v$, 
for the weak bremsstrahlung processes 
$\chi\chi$ $\to$ $W\ell\nu$, including both 
$W^+\ell^-\bar{\nu}_{\ell}$ and $W^-\ell^+\nu_{\ell}$, 
are plotted in Fig.~\ref{fig:sigmav-mchi} as a function of 
$m_{\chi}$ together with 
the contributions of the relevant two-body processes. 
In Fig.~\ref{fig:sigmav-mchi} (a), 
the MSSM parameters are chosen as 
$\tan\beta$ $=$ 2, $\mu$ $=$ 1 TeV, 
$m_{\tilde{q}}$ $=$ 14 TeV, and 
$m_{\tilde{e}}$ $=$ $m_{\tilde{\mu}}$ $=$ 
$m_{\tilde{\tau}L}$ $=$ $m_{\tilde{\tau}R}$ $=$ 240 GeV. 
The trilinear coupling $A_t$ is chosen 
to satisfy the Higgs mass constraint 
$m_h$ $\sim$  125 GeV~\cite{mh-125GeV-ATLAS,mh-125GeV-CMS}, 
although the trilinear coupling for the squarks are 
irrelevant in the present analysis where squark exchange 
diagrams are sufficiently suppressed by the heavy squark masses. 
The bold solid line corresponds to 
the sum of all the leptonic processes, 
$\sum_{\ell}$ $W\ell\bar{\nu}_{\ell}$ $+$ h.c.  
The contributions of $We\nu_e$ and $W\mu\nu_{\mu}$ are identical, 
$(\sigma v)_{We\nu_e}$ $=$ $(\sigma v)_{W\mu\nu_{\mu}}$, 
which are essentially described by the simplified expression
in Appendix~\ref{app:toy-model}. 
On the other hand, the result for 
$W\tau\nu_{\tau}$ is slightly larger 
$(\sigma v)_{W\tau\nu_{\tau}}$ $\sim$
$1.2 \times (\sigma v)_{We\nu_e}$, where the difference 
originates from the left-right mixing for the staus. 
The contributions of the relevant two-body processes 
are shown with thin lines.\footnote{
The p-wave contributions 
$(\sigma v)_2^{\rm p\mathchar`-wave}$ 
$\sim$ $v^2 g^{\prime 4}/(192\pi m_{\chi}^2)$ 
are typically 
smaller than the relevant s-wave contributions for 
the relative velocity $v$ $\sim$ $10^{-3}$ in the galactic halo.} 
The solid, dashed, dot-dashed and dotted lines 
correspond to $\tau^+\tau^-$, 
$t\bar{t}$, $b\bar{b}$ and $W^+W^-$, respectively. 
For a relatively large value of 
$m_{\chi}$ $\lesssim$ $m_{\tilde{\ell}}$, 
the weak bremsstrahlung dominates over the 
two-body contributions. 
The range of $m_{\chi}$ filled with gray 
corresponds to the cosmologically allowed region 
where the relic abundance constraint 
0.11 $<$ $\Omega_{\chi} h^2$ $<$ 0.13 is satisfied. 
The relic abundance is obtained using DarkSUSY~\cite{darksusy}. 
In the present scenario with a bino-like LSP, 
the relic density is typically too large. The allowed region
in Fig.~\ref{fig:sigmav-mchi} (a) appears with the help of 
slepton coannihilations which lead to an effective 
enhancement of the pair annihilation cross section 
for $m_{\chi}$ $\approx$ $m_{\tilde{\ell}}$~\cite{Griest-Seckel,Mizuta-Yamaguchi,ino-coan,stau-coan,NRR3}. 


In Fig.~\ref{fig:sigmav-mchi} (b), 
the soft mass parameter for the right-handed stau 
is taken to be larger than the other slepton mass parameters 
as $m_{\tilde{\tau}R}$ $=$ 480 GeV. 
In this case, the contributions 
of $We\nu_e$ and $W\mu\nu_{\mu}$ are the same as in 
Fig.~\ref{fig:sigmav-mchi} (a), while that of $\tau^+\tau^-$ 
gets suppressed due to the larger $m_{\tilde{\tau}R}$. 
The contribution of $W\tau\nu_{\tau}$ becomes identical to 
those of $We\nu_e$ and $W\mu\nu_{\mu}$, since the effect of 
left-right mixing for the staus, 
$\sim$ 
$m_{\tau}(A_{\tau}-\mu\tan\beta)/(m_{\tilde{\tau}L}^2-m_{\tilde{\tau}R}^2+\delta_{\tau})$, 
is reduced by taking different values for $m_{\tilde{\tau}R}$ and 
$m_{\tilde{\tau}L}$, where 
$\delta_{\tau}$ $\sim$ $-\,0.04 m_Z^2\cos 2\beta$. 
Thus the relative magnitude of weak bremsstrahlung is increased 
compared with the result in Fig.~\ref{fig:sigmav-mchi} (a). 
The weak bremsstrahlung is dominant 
for all values of $m_{\chi}$. 
%


In Fig.~\ref{fig:sigmav-mchi} (c), 
the soft mass parameters for both the left- and right-handed 
stau are taken to be larger than that for the selectron and smuon
as $m_{\tilde{\tau}L}$ $=$ $m_{\tilde{\tau}R}$ $=$ 480 GeV. 
The bold dashed line represents the result for only $W\tau\nu_{\tau}$. 
The contribution of $W\tau\nu_{\tau}$ is reduced due to 
the larger $m_{\tilde{\tau}L}$.


Figure~\ref{fig:sigmav-mchi} (d) is 
the result for $\tan\beta$ $=$ 10, 
$\mu$ $=$ 3 TeV, 
and $m_{\tilde{q}}$ $=$ 12 TeV, 
where the mass parameters for staus are taken to be larger as 
$m_{\tilde{\tau}L}$ $=$ $m_{\tilde{\tau}R}$ $=$ 2 TeV. 
For a large $\tan\beta$, the contribution of 
the $\tau^+\tau^-$ final state is enhanced due to the
larger left-right mixing, 
while $We\nu_e$ and $W\mu\nu_{\mu}$ remain unchanged. 
However, by taking much larger masses for staus 
than in Figs.~\ref{fig:sigmav-mchi} (a)-(c), 
the weak bremsstrahlung can be dominant over 
the two-body processes. 
The process $W\tau\nu_{\tau}$ is suppressed as 
$\sim$ $1/m_{\tilde{\tau}L}^8$. 
The bold dotted line represents the contribution of 
the $Wtb$ final state below the $t\bar{t}$ threshold 
$m_{\chi}$ $<$ $m_t$. 
This proceeds via off-shell top quark effect 
$\chi\chi$ $\to$ $t^*\bar{t}$ 
followed by the decay $t^*$ $\to$ $Wb$~\cite{chen-kam}. 
In the present analysis, we include 
only $t$- and $u$-channel diagrams for  
the $Wtb$ final state as explained in Sec.~\ref{sec:NWA}. 
If we had included the $s$-channel diagrams for $Wtb$, 
the result below the threshold would be smoothly connected to 
the two-body $t\bar{t}$ result. 


We define the ratio $R$ $=$ $(\sigma v)_{3W}/(\sigma v)_2$, 
where $(\sigma v)_{3W}$ is 
the total leptonic $W$-strahlung contribution, and 
$(\sigma v)_2$ $=$ 
$\sum_{f=t, b, \tau}$ $(\sigma v)_{f\bar{f}}$ $+$ $(\sigma v)_{WW}$
is the sum of all the relevant two-body contributions included in the 
present analysis.\footnote{We have not inluded $ZZ$, $HV$ and $HH$ 
final states in the present analysis for simplicity, where $V$ $=$ $W$/$Z$, 
and $H$ represents one of the Higgs bosons.}
The contours of the ratio $R$ are 
plotted in the ($\mu$, $M_2$) plane 
in Fig.~\ref{fig:cont-M2-mu} with bold lines. 
The relavant parameters for Figs.~\ref{fig:cont-M2-mu}
(a)-(d) are taken to be the same as 
in Figs.~\ref{fig:sigmav-mchi} (a)-(d), respectively. 
The contours of $(\sigma v)_{3W}$ 
are shown with thin lines. 
As $M_2$ gets across $2m_t$ $\sim$ 350 GeV, the ratio becomes small 
suddenly due to opening of the $t\bar{t}$ channel. 
The strip filled with gray corresponds to 
the cosmologically allowed region with the correct $\Omega_{\chi}h^2$. 
In Fig.~\ref{fig:cont-M2-mu} (a), 
the $W$-strahlung cross section $(\sigma v)_{3W}$ becomes larger
as $M_2$ increases. 
In the cosmologically allowed range, 
the weak bremsstrahlung can be comparable with the 
two-body processes for 800 GeV $\lesssim$ $\mu$ $\lesssim$ 1100 GeV. 
The area filled with light gray is excluded where the 
lighter stau is lighter than the neutralino. 
The figure includes a small $\mu$ range where 
the LSP is higgsino-like. In this region, 
the $WW$ final state becomes dominant, 
though cosmologically allowed regions exist
for higgsino-like LSP. 
Therefore, the weak bremsstrahlung is negligible 
for a higgsino-like LSP in the present analysis. 
We do not show the result for $\mu$ $<$ 0, since 
the behavior is similar to that for $\mu$ $>$ 0. 


In Figs.~\ref{fig:cont-M2-mu} (b) and (c), 
where the mass parameters for staus are larger, 
the contribution of $W\ell\nu_{\ell}$ is  
similar to that in Fig.~\ref{fig:cont-M2-mu} (a). 
However, the ratio can be significantly larger 
than in Fig.~\ref{fig:cont-M2-mu} (a), since 
the relevant two-body final state, $\tau^+\tau^-$, 
gets smaller than in Fig.~\ref{fig:cont-M2-mu} (a). 
In Fig.~\ref{fig:cont-M2-mu} (d), 
the ratio is further enhanced by taking 
a large value for stau mass parameters as 
$m_{\tilde{\tau}L}$ $=$ $m_{\tilde{\tau}R}$ $=$ 2 TeV
for $\tan\beta$ $=$ 10.


Finally, neutrino spectra at injection in the Sun 
are shown in Fig.~\ref{fig:spect-Enu}. 
The panel~(a) is the result for 
$\tan\beta$ $=$ 2, $M_2$ $=$ 450 GeV, 
$\mu$ $=$ 1 TeV, 
$m_{\tilde{q}}$ $=$ 14 TeV, and 
$m_{\tilde{e}}$ $=$ $m_{\tilde{\mu}}$ $=$ 
$m_{\tilde{\tau}L}$ $=$ $m_{\tilde{\tau}R}$ $=$ 240 GeV. 
The bold solid line corresponds to 
the primary neutrino spectrum of weak bremsstrahlung 
including all the flavors and the charge conjugated states. 
The bold dashed and dotted lines are the results for 
the secondary neutrinos from the tau lepton and the $W$ boson, 
respectively, produced via weak bremsstrahlung. 
The contributions of two-body processes are drawn with  
thin lines. 
The neutralino mass in this case is $m_{\chi}$ $\approx$ 221.9 GeV. 
One can see that 
the primary neutrino from weak bremsstrahlung can give 
a significant contribution particularly in the 
high energy range 
$0.8 m_{\chi}$ $\lesssim$ $E_{\nu}$ $\lesssim$ $m_{\chi}$. 
The result for 
$m_{\tilde{\tau}L}$ $=$ $m_{\tilde{\tau}R}$ $=$ 480 GeV 
is shown in the panel~(b). 
In this case, the weak bremsstrahlung is dominant for 
the wide range of $E_{\nu}$, since 
the $\tau^+\tau^-$ final state is suppressed by the 
large $m_{\tilde{\tau}L}$ and $m_{\tilde{\tau}R}$. 
It is notable that 
primary neutrino contributions of weak bremsstrahlung are nearly 
flavor independent for a common slepton mass, 
while the largest contribution 
among the two-body process, $\tau^+\tau^-$, 
produces mainly tau neutrinos in the Sun. 
Hence $W\ell \nu$ can strongly affect the flavor contents 
of energetic neutrinos~\cite{bell-neutrino-osc}.

Even when the three-body final state $W\ell \nu$ is not the dominant channel
in the total cross section, 
the neutrino spectrum at high energies, 
$E_{\nu}$ $\lesssim$ $m_{\chi}$, can still be dominated by $W\ell \nu$. 
The energetic neutrinos mainly originate from the 
internal bremsstrahlung, diagram C in Fig.~\ref{fig:diagrams}. 
Indeed, in Fig.~\ref{fig:spect-Enu}(a), 
the total cross section of the three-body process is slightly smaller than 
the sum of the two-body results. 
Nevertheless, the three-body process is dominant in the high energy region in 
Fig.~\ref{fig:spect-Enu}(a).

It must be kept in mind that one should take into account 
the usual bremsstrahlung effect $(\sigma v)_{3\gamma}$ as well 
to evaluate the total three-body cross section. 
However, as far as a neutrino flux is concerned, 
the weak bremsstrahlung is expected to give the major contribution. 
Also, $Z$-boson strahlung processes, e.g., 
$\chi\chi$ $\to$ $Z\ell^+\ell^-$, should be included to discuss 
the total contributions from weak bremsstrahlung.

%
\section{Conclusions}
\label{sec:conclusions}
%

We have examined the effects of electroweak bremsstrahlung 
on the bino-like neutralino dark matter pair annihilations 
in the MSSM. 
We have calculated the nonrelativistic pair annihilation 
cross sections and neutrino spectra via $W$-strahlung, 
$\chi\chi$ $\to$ $W\ell\bar{\nu}$. 
It has been shown that the electroweak bremsstrahlung 
can give a dominant contribution to the cross section 
for some parameter regions which include cosmologically 
allowed ranges with the observed relic abundance. 
It has been found that 
the weak bremsstrahlung processes can give characteristic 
signals in the neutrino spectrum at injection in the Sun.

In the present analyses, we considered extremely heavy
squarks. When the squark masses are comparable with 
the slepton masses, the weak bremsstrahlung typically gives 
only a subdominant contribution for $m_{\chi}$ $>$ $m_t$
due to the unsuppressed $t\bar{t}$ final state. 
Below the $t\bar{t}$ threshold, the $Wtb$ final state 
can be relevant as shown in Ref.~\cite{chen-kam}.

%
% ... 
%
%
\section*{Acknowledgments}
This work was supported in part by a CST grant for Gakujyutsusyo, 
Nihon University. 
The authors are very grateful to S.~Naka, S.~Deguchi and A.~Miwa 
for useful discussions and comments.

%
\appendix 

%
\section{Unbroken SU(2) limit in the slepton sector} 
\label{app:toy-model}

In this section, we provide a simple expression for 
the unbroken $SU(2)_L$ limit in the slepton sector, 
taking the common slepton masses 
$m_{\tilde{\nu}}$ $=$ $m_{\tilde{\ell}I}$ $\equiv$ 
$\tilde{m}$, 
and the common left-handed slepton coupling constants 
$C^{(\nu)}_L$ $=$ $C^{(\ell)}_{L1}$ $\equiv$ $C_L$
with $C^{(\ell)}_{L2}$ $=$ $C^{(\ell)}_{RI}$ $=$ 0, 
and keeping $m_W$ $\neq$ 0. 
This corresponds to the choices 
$N_{12}$ $=$ 0, $h_{\ell}$ $=$ 0, 
$\big( \widetilde{V}_{\ell} \big)_{11}$ $=$ 1 and 
$\big( \widetilde{V}_{\ell} \big)_{12}$ $=$ 0 
in Eq.~(\ref{eqn:C-fL-I}). 
In this limit, the neutralino is a pure bino, and 
only the left-handed sleptons contribute. 
Writing the matrix element ${\cal M}$ $=$ 
${\cal M}_{\rm A}$ $+$${\cal M}_{\rm B}$ $+$${\cal M}_{\rm C}$ 
as ${\cal M}$ $=$ $\varepsilon_3^{\mu} {\cal M}_{\mu}$, 
it is confirmed that the Ward identity 
$p_3^{\mu} {\cal M}_{\mu}$ $=$ 0 is satisfied 
using $\!\not{\! p}_1 u_{\ell}$ 
$=$ $ \!\not{\! p}_2 v_{\nu}$ $=$ 0~\cite{bell-etal-revisited}. 
This implies that the longitudinal polarization of the 
$W$ boson does not contribute to the s-wave amplitude 
in this limit.

The differential cross section for $\chi\chi$ $\to$ $W\ell\bar{\nu}$ 
can be greatly simplified as 
\begin{eqnarray}
\frac{d^2 (\sigma v)_{W\ell\bar{\nu}}}{dE_W \, dE_{\nu}}
 & = & 
\frac{g^2 C_L^4}{4096\pi^3 m_{\chi}^2} 
\frac{1}{(t_1-\tilde{m}^2)^2} \frac{1}{(t_2-\tilde{m}^2)^2} 
y \left( x^2 + z^2 -8 m_{\chi}^2 m_W^2 \right). 
\label{eqn:d2svdEdE-lepton}
\end{eqnarray}
Integrating over $dE_{\nu}$ and $dE_W$ analytically, 
Eq.~(\ref{eqn:d2svdEdE-lepton}) reduces to the 
cross section found in Ref.~\cite{bell-etal-revisited}. 
Note that the cross section is severely suppressed as 
$\sim$ $1/\tilde{m}^8$ in the heavy slepton limit.

In the numerical analyses in Sec.~\ref{sec:numerical-results},  
the $SU(2)_L$ breaking effects lead to 
$m_{\tilde{\nu}}^2-m_{\tilde{\ell}1}^2$ $\sim$ 
$m_Z^2 \cos 2\beta \cos^2 \theta_W$ 
without the left-right mixing for the sleptons.  
In the presence of the mass splitting 
$m_{\tilde{\nu}}-m_{\tilde{\ell}1}$ $\neq$ 0, 
a longitudinal $W$-boson emission 
 $\chi\chi$ $\to$ $W_L \ell \bar{\nu}_{\ell}$ 
can enhance the cross section~\cite{antiproton}.\footnote{Neglecting the slepton mixing, but still taking the mass splitting into account, our result agrees with the one found in Ref.~\cite{antiproton}.}

%
\section{Two-body processes}
\label{app:2-body}

In this appendix, the s-wave cross sections for 
the relevant two-body processes are summarized 
for convenience~\cite{Jungman-etal}.

The s-wave cross section for the fermion pair production 
$\chi\chi \to f\bar{f}$ is given by 
\begin{eqnarray}
(\sigma v)_{f\bar{f}}
 & = & 
\frac{N_c}{2\pi}
\sqrt{1-\frac{m_f^2}{m_{\chi}^2}} 
\left| F_A + F_Z + \sum_{I=1}^2 F_{\tilde{f}I}  \right|^2, 
\label{eqn:sv-ffbar}
\end{eqnarray}
where 
\begin{eqnarray}
F_A  & = & 
\frac{C_P^{ffA}C_P^{\chi\chi A}}{4m_{\chi}^2-m_A^2+ i\Gamma_A m_A}
  m_{\chi}, \nonumber \\
F_Z  & = & 
\frac{C_A^{ffZ}C_A^{\chi\chi Z}}{4m_{\chi}^2-m_Z^2+ i\Gamma_Z m_Z}
  \frac{m_f\left(4m_{\chi}^2-m_Z^2\right)}{m_Z^2}, 
\label{eqn:2body-Amp} \\
F_{\tilde{f}I}  & = & \frac{1}{4} \cdot 
\frac{m_f \left[ (C^{(f)}_{LI})^2 + (C^{(f)}_{RI})^2 \right]
+ 2m_{\chi}C^{(f)}_{LI} C^{(f)}_{RI}}{m_f^2-m_{\chi}^2-m_{\tilde{f}I}^2}. 
\nonumber 
\end{eqnarray}
The quantities $F_A$, $F_Z$ and $F_{\tilde{f}I}$ represent 
the amplitude of $s$-channel pseudoscalar Higgs boson ($A$) 
exchange, $s$-channel $Z$-boson exchange, and $t$- and $u$-channel
sfermion exchange, respectively. 
The constant $N_c$ is the color factor: 
$N_c$ $=$ 3 for quark pairs, and $N_c$ $=$ 1 for leptons. 
The coupling constants 
$C_P^{ffA}$ and $C_P^{\chi\chi A}$ describe the interaction of 
the pseudoscalar Higgs $A$ with bilinears $\bar{f}i\gamma_5f$ 
and $\bar{\chi}i\gamma_5\chi$, respectively. 
The coupling constants 
$C_A^{ffZ}$ and $C_A^{\chi\chi Z}$ determine 
the axial vector interaction of 
the $Z$ boson with bilinears $\bar{f}\gamma_{\mu}\gamma_5f$ 
and $\bar{\chi}\gamma_{\mu}\gamma_5\chi$, respectively. 
The expressions for these coupling constants 
can be found in Ref.~\cite{NRR2}. 
The decay widths for the $Z$ boson and the pseudoscalar Higgs, 
$\Gamma_Z$ and $\Gamma_A$, are taken into account. 
Note that the contribution of the sfermion exchange, $F_{\tilde{f}I}$,  
includes a factor of the fermion mass $m_f$, since 
$C^{(f)}_{RI}$ ($C^{(f)}_{LI}$) is proportional to $m_f$ for 
$I$ $=$ 1 ($I$ $=$ 2).

The s-wave cross section for the $W$-boson pair production
$\chi \chi \to W^+ W^-$ via chargino exchange diagrams 
is given by 
\begin{eqnarray}
(\sigma v)_{WW}
& = & 
\frac{1}{2\pi} \sqrt{1-\frac{m_W^2}{m_{\chi}^2}}
(m_{\chi}^2-m_W^2) 
\left|  \sum_{p=1,2}
 \frac{
\left(C_{V}^{\chi_{p}^{+}\chi W^{-}}\right)^2 + \left(C_{A}^{\chi_{p}^{+}\chi W^{-}}\right)^2}{m_{\chi}^2+m_{\chi^+_p}^2-m_W^2} \right|^2, 
\label{eqn:sv-WW}
\end{eqnarray}
where $m_{\chi^+_p}$ denotes the chargino mass ($p$ $=$ 1, 2). 
The coupling constants 
$C_{V}^{\chi_{p}^{+}\chi W^{-}}$ and 
$C_{A}^{\chi_{p}^{+}\chi W^{-}}$ 
defined in Ref.~\cite{NRR2} 
describe the neutralino-chargino-W vector/axial-vector interactions.

Neutrino spectra for the two-body processes can be obtained from 
Eq.~(\ref{eqn:spectrum-3body}) by replacing the primary 
spectrum with the delta function distribution. 
For instance, the neutrino spectrum via $\chi\chi$ $\to$ $W^+W^-$ 
can be written as 
\begin{eqnarray}
\left. \frac{d(\sigma v)_{WW}}{dE_{\nu}} \right|_{{\rm from} \, W}
 & = & 
(\sigma v)_{WW} 
\left[ \left(\frac{dN_{\nu}}{dE_{\nu}}\right)_W 
                       (m_{\chi}, E_{\nu}) \right], 
\end{eqnarray}
by replacing $d (\sigma v)_{W\ell\bar{\nu}}/dE_W$ 
in Eq.~(\ref{eqn:spectrum-3body}) 
with  $(\sigma v)_{WW} \delta (E_W-m_{\chi})$.


%

\begin{figure}[p]
\includegraphics[width=7cm]{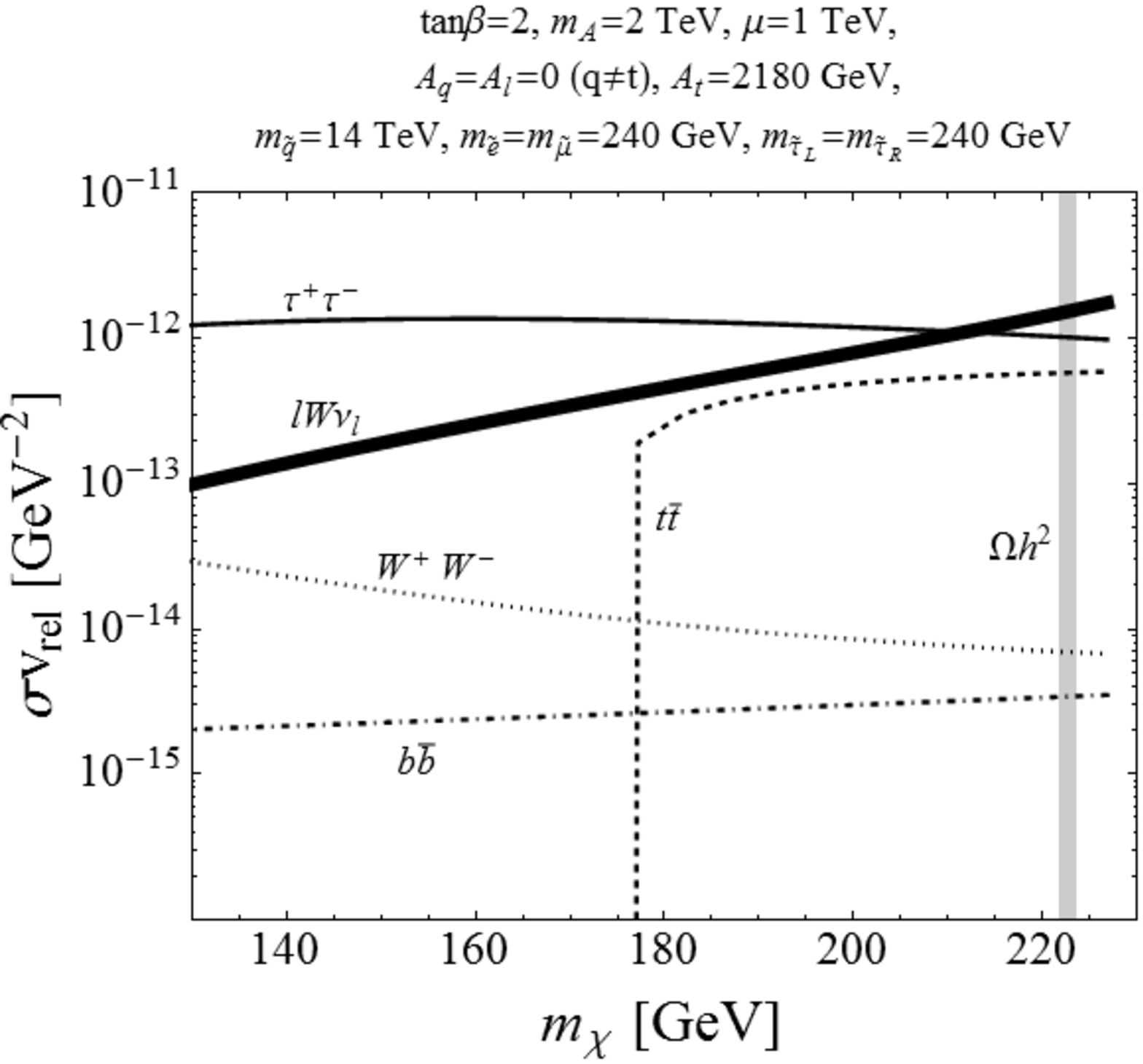}
\hspace{5mm}
\includegraphics[width=7cm]{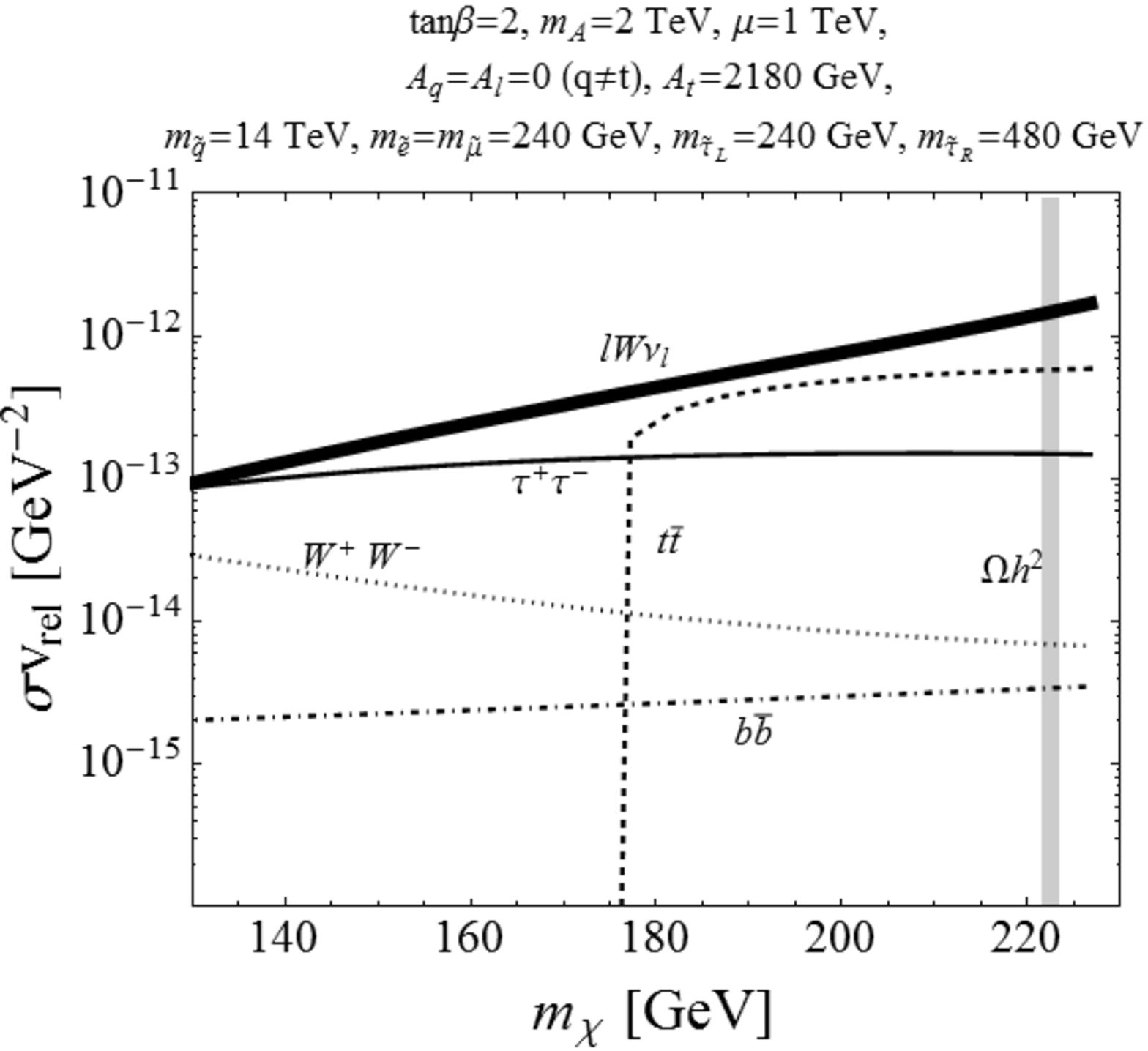} \vspace{1cm} \\
\includegraphics[width=7cm]{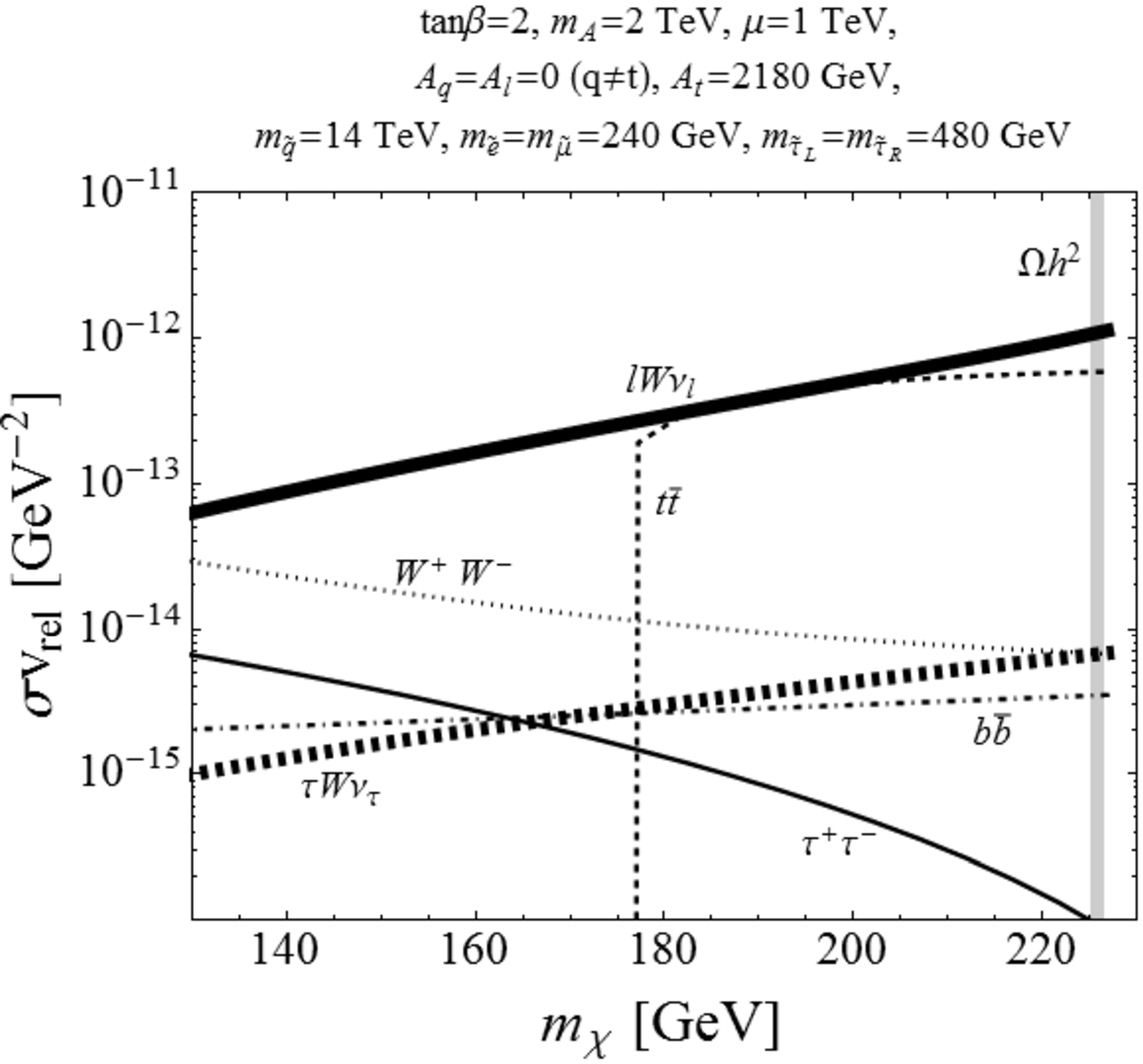}
\hspace{5mm}
\includegraphics[width=7cm]{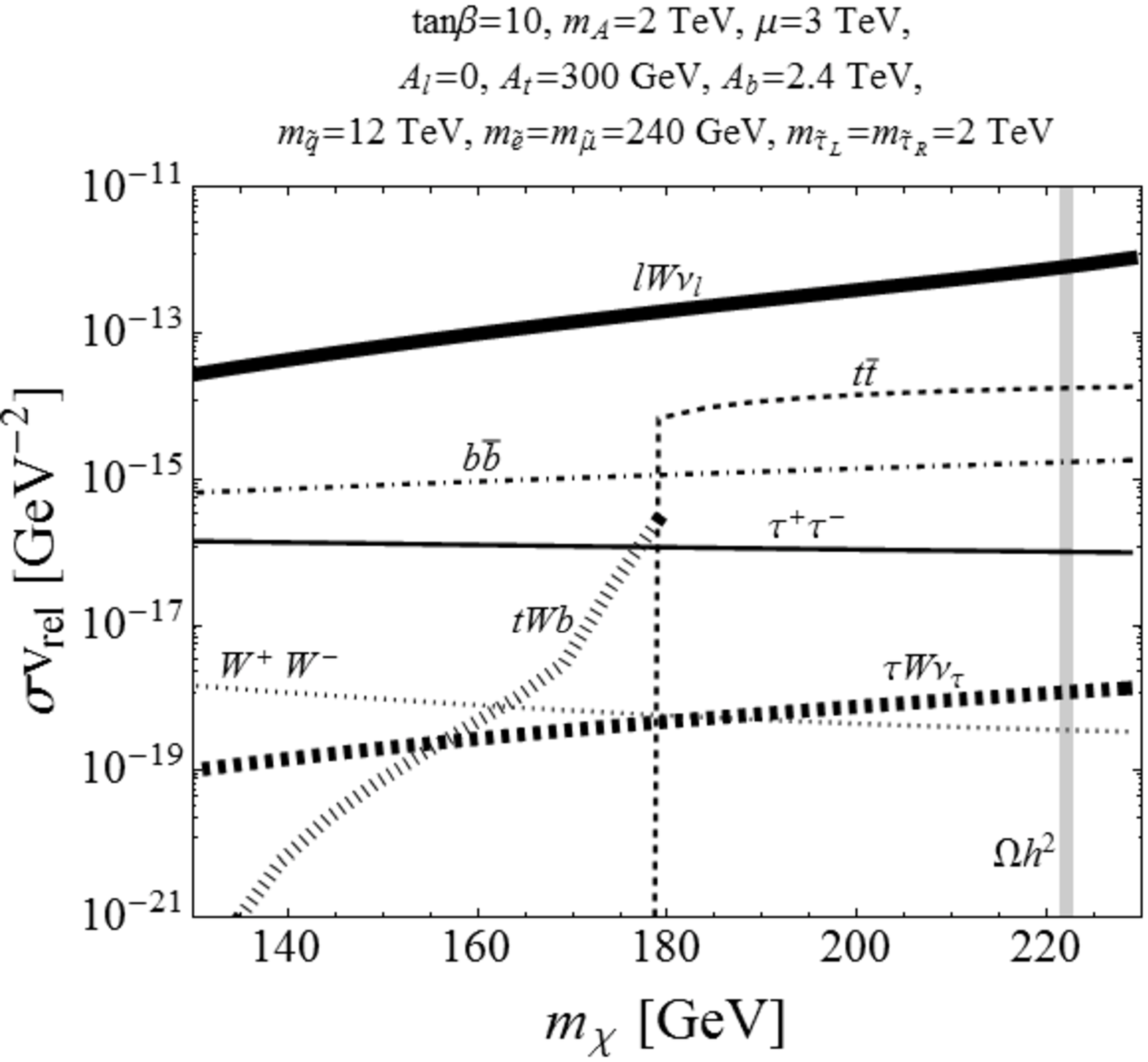} 
\vspace{-10.3cm} \\ 
\unitlength 1mm
\begin{picture}(100,100)(18,0)
\put(0,70){(a)}
\put(77,70){(b)}
\put(0,0){(c)}
\put(77,0){(d)} 
\end{picture}
\caption[cap:sigmav-mchi]{
\baselineskip 6mm 
(a) The cross section times the relative velocity, $\sigma v$, 
for the weak bremsstrahlung processes 
$\chi\chi$ $\to$ $\sum_{\ell}$ $W\ell\nu$ (bold solid line), 
including both 
$W^+\ell^-\bar{\nu}_{\ell}$ and $W^-\ell^+\nu_{\ell}$, 
as a function of $m_{\chi}$. 
In panels (b), (c) and (d), 
the result for only $W\tau\nu_{\tau}$ is shown with 
the bold dashed line. 
The bold dotted line represents the contribution of $Wtb$ 
evaluated with only $t$- and $u$-channel diagrams. 
The contributions of the relevant two-body processes are shown
with thin lines. 
The solid, dashed, dot-dashed and dotted lines 
correspond to $\tau^+\tau^-$, 
$t\bar{t}$, $b\bar{b}$ and $W^+W^-$, respectively. 
The range of $m_{\chi}$ filled with gray 
corresponds to the cosmologically allowed region 
where the relic abundance constraint 
0.11 $<$ $\Omega_{\chi} h^2$ $<$ 0.13 is satisfied. 
}
\label{fig:sigmav-mchi}
\end{figure}
%

\begin{figure}[p]
\includegraphics[width=7cm]{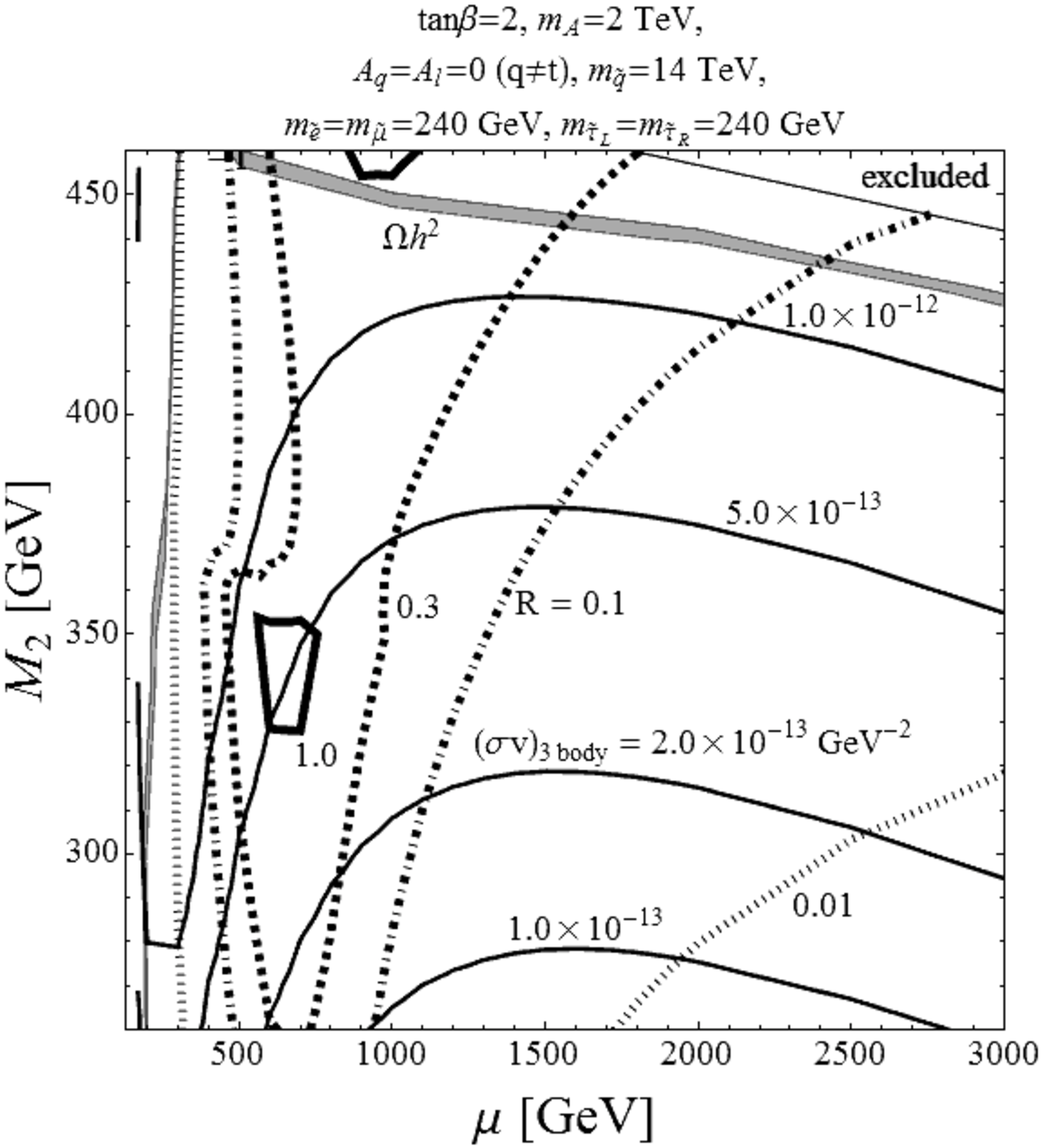} 
\hspace{5mm}
\includegraphics[width=7cm]{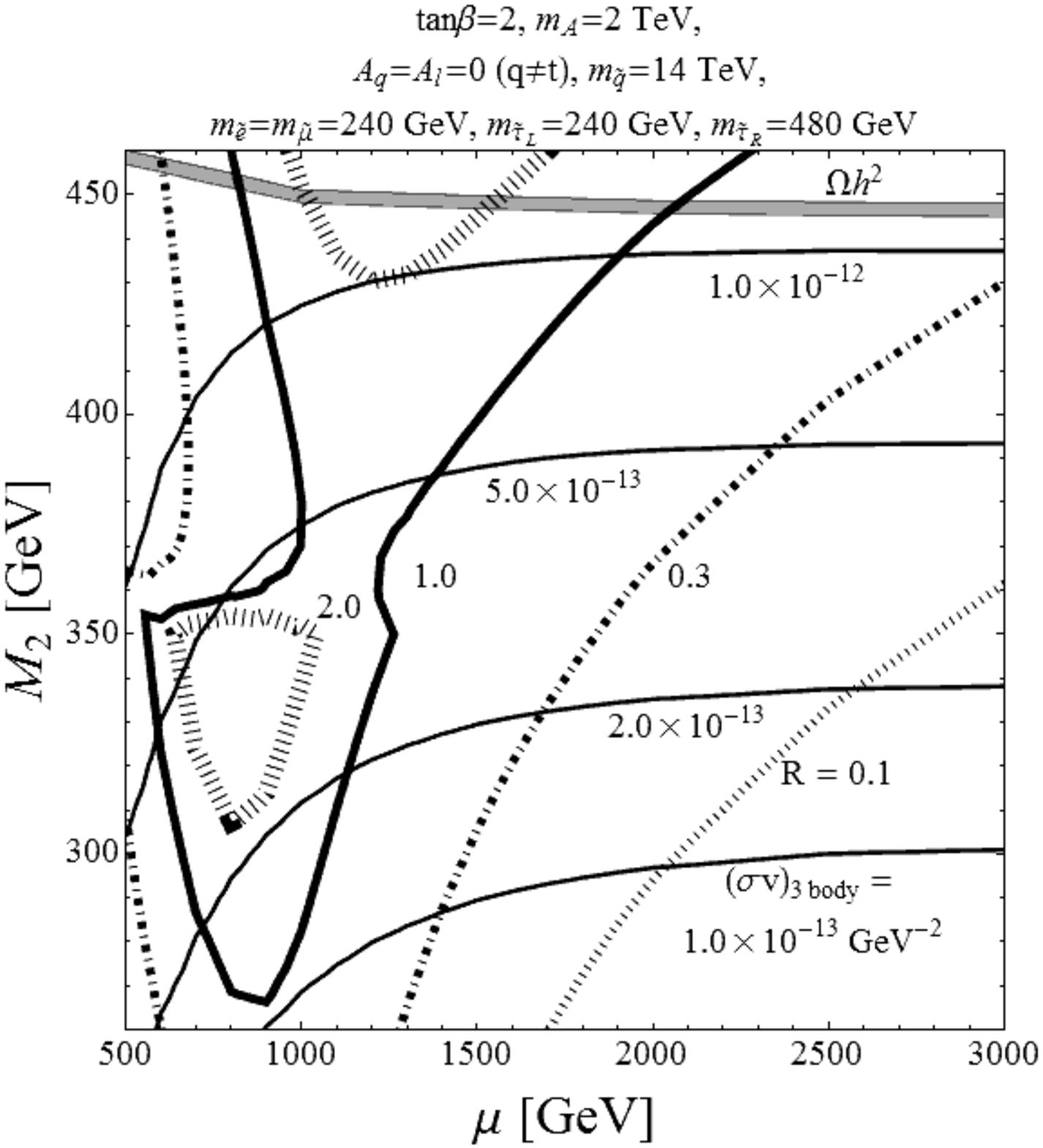}
\vspace{1cm} \\
\includegraphics[width=7cm]{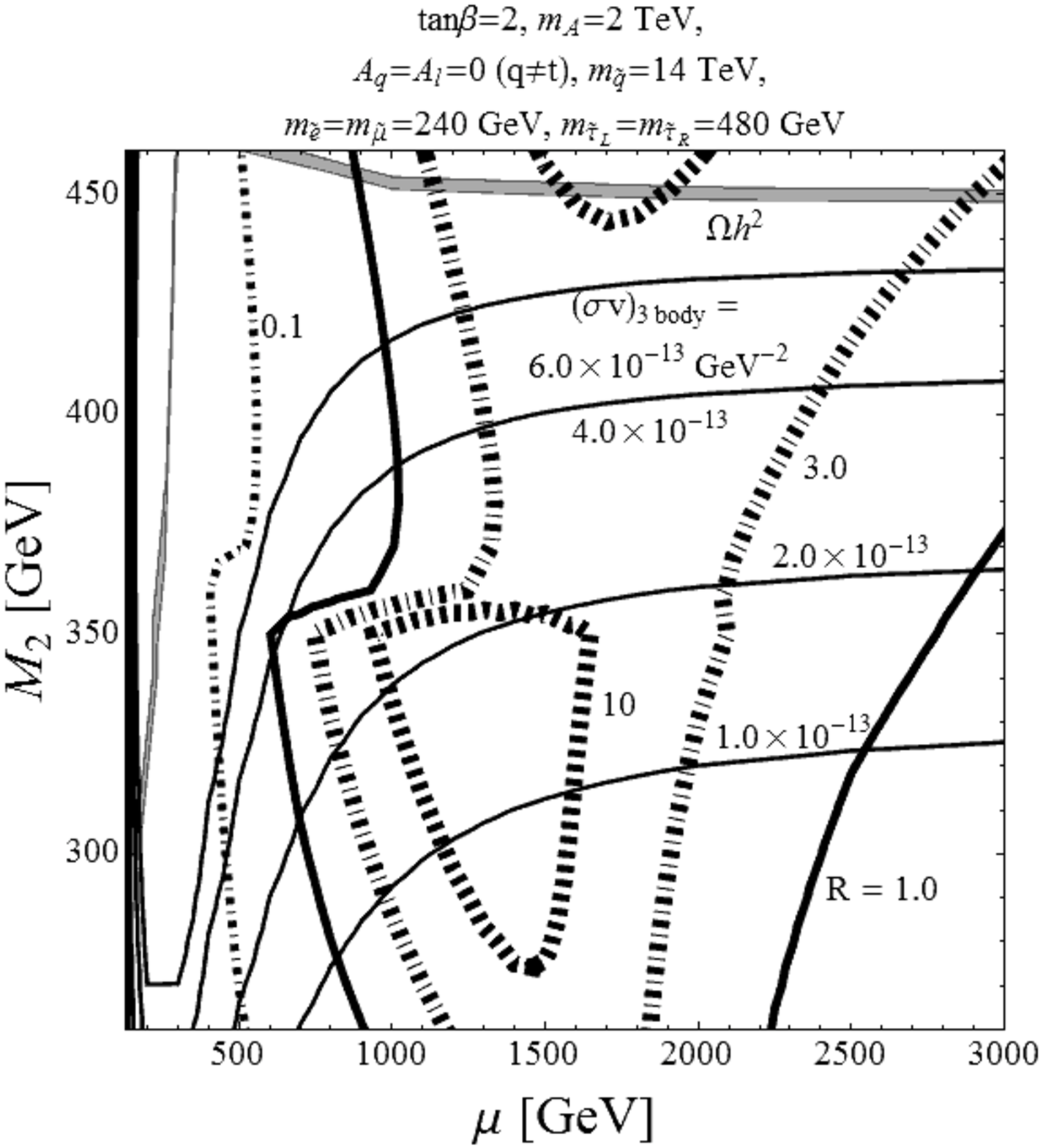}
\hspace{5mm}
\includegraphics[width=7cm]{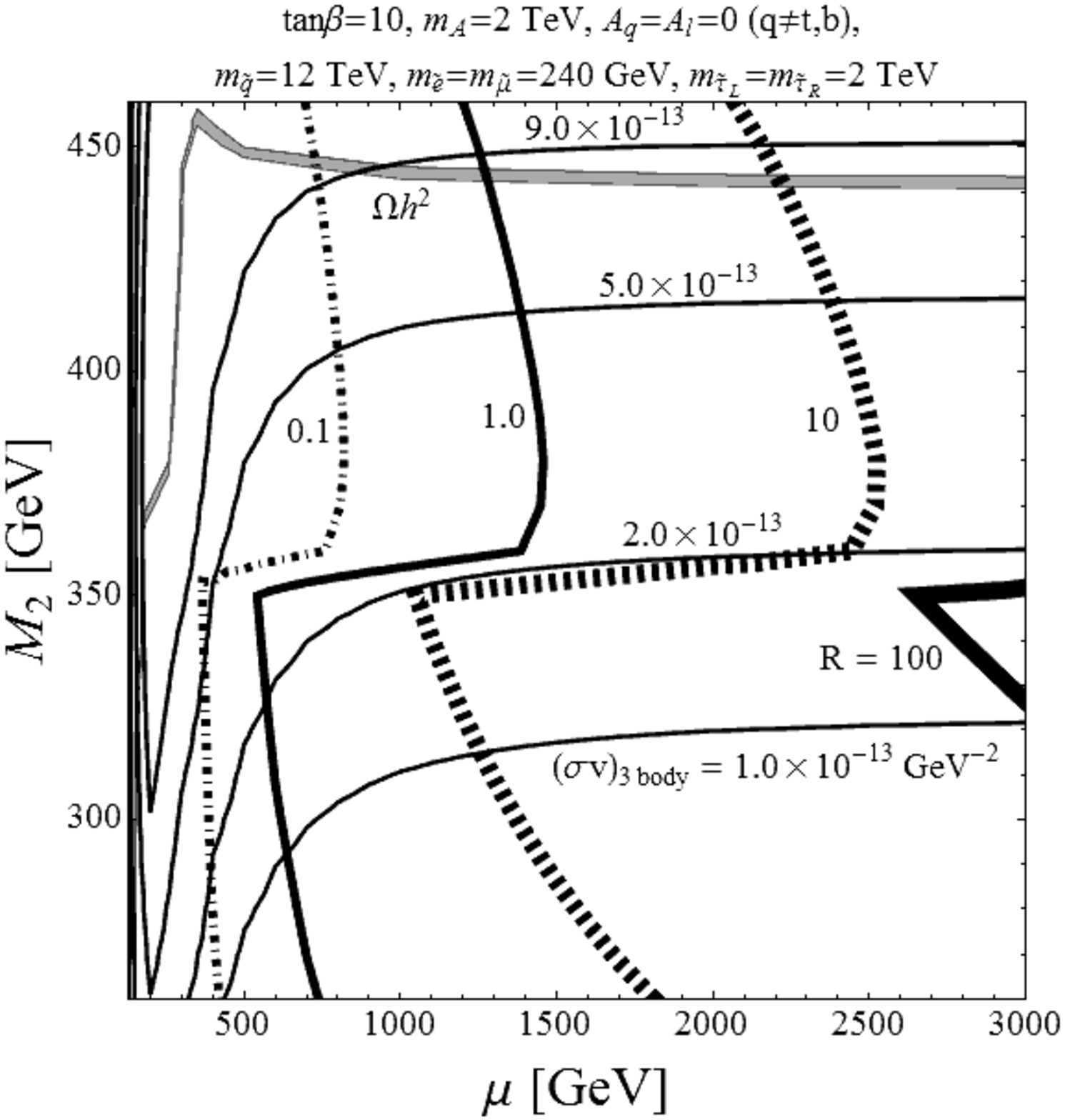} 
\vspace{-10.1cm} \\ 
\unitlength 1mm
\begin{picture}(100,100)(22,0)
\put(0,85){(a)}
\put(77,85){(b)}
\put(0,0){(c)}
\put(77,0){(d)} 
\end{picture}
\caption[cap:cont-M2-mu]{
\baselineskip 6mm 
Contours of the total $W$-strahlung contribution $(\sigma v)_{3W}$
(thin solid lines). 
Contours of 
the ratio $R$ $=$ $(\sigma v)_{3W}/(\sigma v)_2$
are shown in bold lines, 
where $(\sigma v)_2$ 
is the sum of all the two-body contributions. 
The strip filled with gray corresponds to 
the cosmologically allowed region with the correct $\Omega_{\chi}h^2$. 
}
\label{fig:cont-M2-mu}
\end{figure}
%

\begin{figure}[p]
\includegraphics[width=7cm]{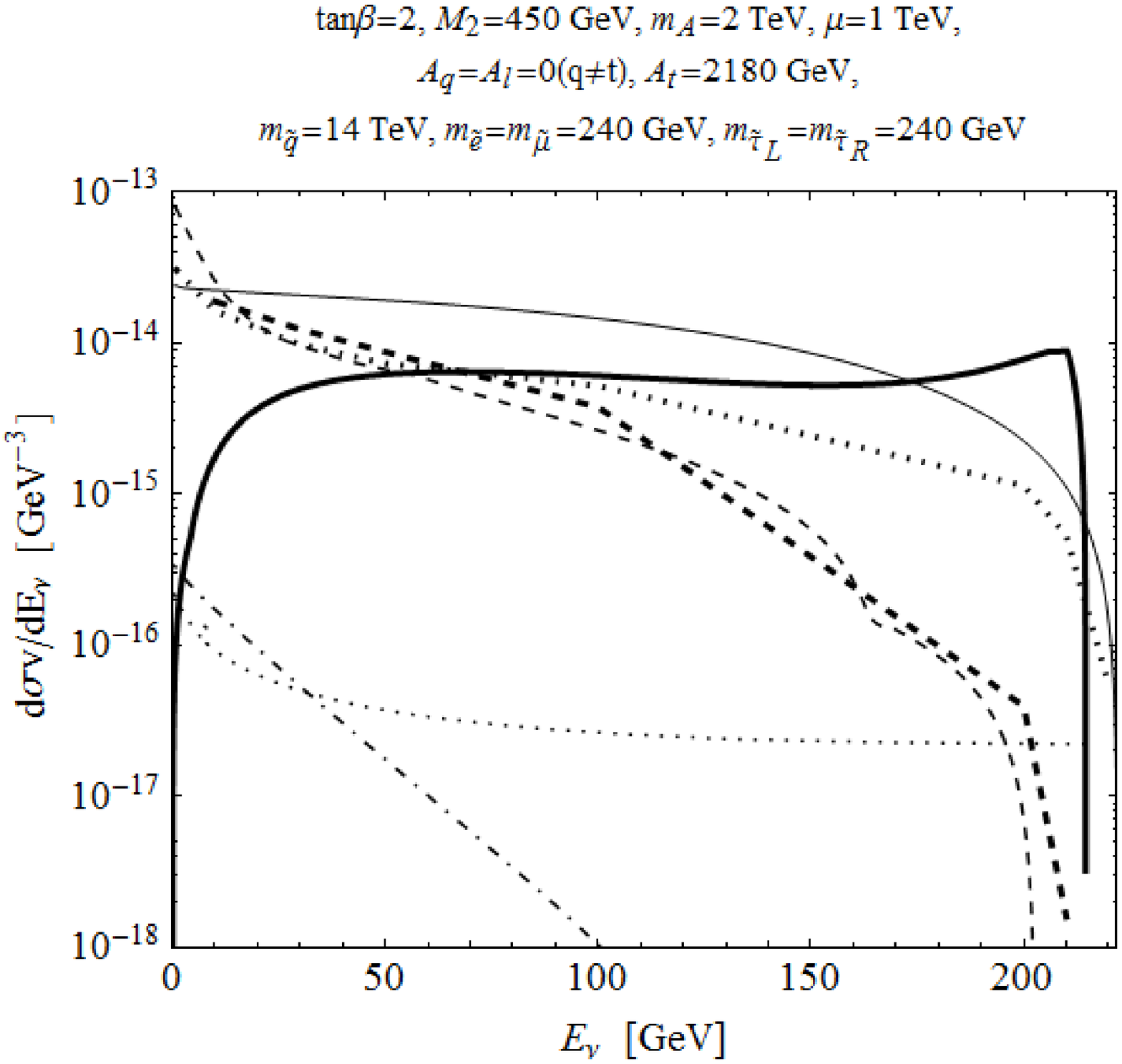}
\hspace{5mm}
\includegraphics[width=7cm]{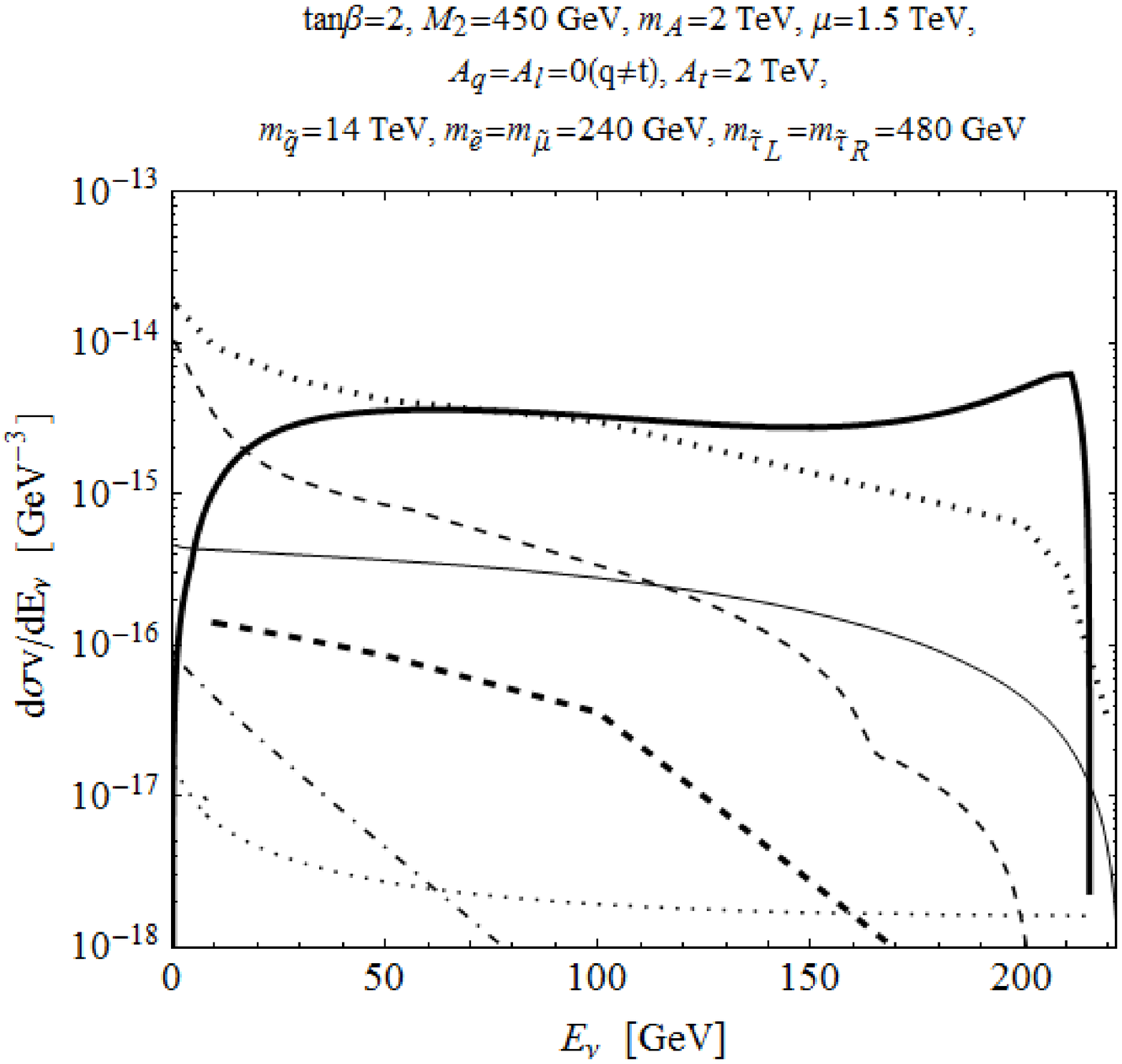}
\vspace{-5cm} \\ 
\unitlength 1mm
\begin{picture}(100,50)(22,0)
\put(0,0){(a)}
\put(77,0){(b)} 
\end{picture}
\caption[cap:spect-Enu]{
\baselineskip 6mm 
Neutrino spectra at injection from the center of the Sun. 
The bold solid line corresponds to 
the primary neutrino spectrum of weak bremsstrahlung 
including all the flavors and the charge-conjugated states. 
The bold dashed and dotted lines are the results for 
the secondary neutrinos from the tau lepton and the $W$ boson, 
respectively, produced via weak bremsstrahlung. 
The contributions of two-body processes are drawn with  
thin lines. 
}
\label{fig:spect-Enu} 
\end{figure}
%

%
%
%

\begin{thebibliography}{99}
%
\baselineskip 6mm 
%
\bibitem{Kolb-Turner}
E. Kolb and M. Turner, {\em The Early Universe}, Addison-Wesley (1990).
%
%
\bibitem{WMAP7}
P.A.R.~Ade {\it et al.}, [Planck Collaboration], arXiv:1303.5076. 
%
%
\bibitem{Jungman-etal}
For reviews on supersymmetric dark matter, see for instance, 
G. Jungman, M. Kamionkowski and K. Griest, \Journal{\PR}{267}{195}{1996}.
%
%
\bibitem{dm-recent-review}
For recent comprehensive reviews on particle dark matter, see for instance, 
G. Bertone (ed.), ``Particle dark matter: Observations, models and searches,'' Cambridge U.P., UK (2010). 
%
%
\bibitem{MSSM}
For a review on the MSSM, see for instance, 
H.P.~Nilles, \Journal{\PR}{110}{1}{1984}. 
%
%
\bibitem{haber-kane} 
H.E.~Haber and G.L.~Kane, \Journal{\PR}{117}{75}{1985}; 
J.F.~Gunion and H.E.~Haber, \Journal{\NPB}{272}{1}{1986}. 
%
%
\bibitem{ffbar}
H.~Goldberg, \Journal{\PRL}{50}{1419}{1983}; 
J.R.~Ellis, J.S.~Hagelin, D.V.~Nanopoulos, K.A.~Olive and M.~Srednicki, 
\Journal{\NPB}{238}{453}{1984}. 
%

%
\bibitem{bergstrom}
L.~Bergstrom, \Journal{\PLB}{225}{372}{1989};  
%
R.~Flores, K.A.~Olive and S.~Rudaz, 
\Journal{\PLB}{232}{377}{1989}. 
%
\bibitem{f-fbar-gamma} 
T.~Bringmann, L.~Bergstrom and J.~Edsjo, 
\Journal{\JHEP}{0801}{049}{2008} [arXiv:0710.3169]; 
%
L. Bergstrom, T. Bringmann and J. Edsjo, 
\Journal{\PRD}{78}{103520}{2008} [arXiv:0808.3725]. 


%
\bibitem{bell-etal-1st-visit}
N.F.~Bell, J.B.~Dent, T.D.~Jacques, L.M.~Krauss and T.J. Weiler, 
\Journal{\PRD}{83}{013001}{2011} [arXiv:1009.2584]. 
%
\bibitem{bell-etal-revisited}
N.F.~Bell, J.B.~Dent, A.J.~Galea, T.D.~Jacques, L.M.~Krauss and T.J. Weiler, 
\Journal{\PLB}{706}{6}{2011} [arXiv:1104.3823]. 
%
\bibitem{gamma-rays-wb} 
V.~Barger, Y.~Gao, W.Y.~Keung and D.~Marfatia, 
\Journal{\PRD}{80}{063537}{2009} [arXiv:0906.3009]; 
Barger, W. -Y. Keung and D. Marfatia, 
\Journal{\PLB}{707}{385}{2012} [arXiv:1111.4523]. 


%
\bibitem{bell-cosmic-rays} 
N.F.~Bell, J.B.~Dent, T.D.~Jacques and T.J. Weiler, 
\Journal{\PRD}{84}{103517}{2011} [arXiv:1101.3357]. 
%
\bibitem{bell-neutrino-osc} 
N.F.~Bell, A.J.~Brennan and T.D.~Jacques, 
\Journal{\JCAP}{1210}{045}{2012} [arXiv:1206.2977]. 
%
\bibitem{fukushima-etal}
K. Fukushima, Y. Gao, J. Kumar and D. Marfatia, 
\Journal{\PRD}{86}{076014}{2012} [arXiv:1208.1010]. 
%
\bibitem{brems-wino-like}
P.~Ciafaloni, M.~Cirelli, D.~Comelli, A.De~Simone, A.~Riotto and A.~Urbano, 
\Journal{\JCAP}{1110}{034}{2011} [arXiv:1107.4453]; 
P.~Ciafaloni, D.~Comelli, A.~De~Simone, A.~Riotto and A.~Urbano, 
\Journal{\JCAP}{1206}{016}{2012} [arXiv:1202.0692]. 
%
\bibitem{kumar-etal}
J. Kumar and P. Sandick. arXiv:1303.2384. 
%
\bibitem{baro-etal}
N. Baro, M. Beneke, M. Kramer, and M. Rummel, PoS IDM2010, 059 (2011). 
%
\bibitem{antiproton}
M. Garny, A. Ibarra and S. Vogl, 
\Journal{\JCAP}{1107}{028}{2011} [arXiv:1105.5367]; 
%
M. Garny, A. Ibarra and S. Vogl,
\Journal{\JCAP}{1204}{033}{2012} [arXiv:1112.5155]. 
%

\bibitem{chen-kam}
X. Chen and M. Kamionkowski, 
\Journal{\JHEP}{9807}{001}{1998} [hep-ph/9805383]. 
%

%
\bibitem{ew-oh2}
P.~Ciafaloni, D.~Comelli, A.~De Simone, E.~Morgante, A.~Riotto 
and A.~Urbano, 
arXiv:1305.6391. 


\bibitem{weak-brems-misc}
P.~Ciafaloni, M.~Cirelli, D.~Comelli, A.~De~Simone, 
A.~Riotto and A.~Urbano, 
\Journal{\JCAP}{1106}{018}{2011} [arXiv:1104.2996]; 
%
M.~Kachelriess, P.D.~Serpico and M.Aa.~Solberg, 
\Journal{\PRD}{80}{123533}{2009} [arXiv:0911.0001]; 
%
J.~Kearney and A.~Pierce, 
\Journal{\PRD}{86}{043527}{2012} [arXiv:1202.0284]. 
%


\bibitem{NRR2}
T. Nihei, L. Roszkowski and R. Ruiz de Austri, 
\Journal{\JHEP}{0203}{031}{2002}.

%
\bibitem{secondary-neutrino}
M.~Cirelli, N.~Fornengo, T.~Montaruli, I.A.~Sokalski, A.~Strumia 
and F.~Vissani, 
\Journal{\NPB}{727}{99}{2005}, Erratum-ibid. {\bf B 790} (2008) 338. 
%
%
\bibitem{pythia}
T. Sj\"{o}strand {\it et.al.}, 
{\em Comput. Phys. Commun} {\bf 135}, 238-259 (2001). 
%

%
\bibitem{top-decay}
J. Beringer {\it et al.} [Particle Data Group], 
\Journal{\PRD}{86}{010001}{2012}. 



\bibitem{LHC-susy-constraint} 
ATLAS Collaboration, 
\Journal{\PLB}{710}{67}{2012} [arXiv:1109.6572], 
CMS Collaboration, 
\Journal{\PRL}{107}{221804}{2011} [arXiv:1109.2352]. 


\bibitem{mh-125GeV-ATLAS}
ATLAS Collaboration, G. Aad {\it et al.}, 
\Journal{\PLB}{716}{1}{2012} [arXiv:1207.7214]. 
%
\bibitem{mh-125GeV-CMS}
CMS Collaboration, S. Chatrchyan {\it et al.}, 
\Journal{\PLB}{716}{30}{2012} [arXiv:1207.7235].

%
\bibitem{darksusy}
P.~Gondolo, J.~Edsj\"{o}, P.~Ullio, L.~Bergstr\"{o}m, 
M.~Schelke and E.A.~Baltz, 
\Journal{\JCAP}{07}{008}{2004} [astro-ph/0406204]. 
%


%
\bibitem{Griest-Seckel}
K. Griest and D. Seckel, \Journal{\PRD}{43}{3191}{1991}.
%
\bibitem{Mizuta-Yamaguchi}
S. Mizuta and M. Yamaguchi, \Journal{\PLB}{298}{120}{1993}.
%
\bibitem{ino-coan}
J. Edsj\"o and P. Gondolo, \Journal{\PRD}{56}{1879}{1997}.
%
\bibitem{stau-coan}
J.R. Ellis, T. Falk and K.A. Olive, \Journal{\PLB}{444}{367}{1998}; 
J.R. Ellis, T. Falk, K.A. Olive and M. Srednicki, 
\Journal{\APP}{13}{181}{2000}.
%
\bibitem{NRR3}
T. Nihei, L. Roszkowski and R. Ruiz de Austri, 
\Journal{\JHEP}{0207}{024}{2002}.
%


%
\end{thebibliography}
\end{document}